\documentclass{aa}  

\usepackage{graphicx}
\usepackage{txfonts}
\usepackage{color}
\usepackage{hyperref}

\usepackage[below]{placeins}
\begin{document}

   \title{TIC 433545934: The first 2+2 type doubly eclipsing binary with extra, mutual eclipses}

   \author{T. Borkovits\inst{1,2,3}  
	  \and
	  S. A. Rappaport\inst{1,4}
          \and
	  P. Zasche\inst{5}
	  \and
	  M. Ma\v{s}ek\inst{6}
	  \and
	  H. Ku\v{c}\'akov\'a\inst{5,7,8}
	  \and
	  T. Mitnyan\inst{1,9}
	  \and
	  V. B. Kostov\inst{10,11}
	  \and
	  B. P. Powell\inst{10}
          }

   \institute{HUN--REN--SZTE Stellar Astrophysics Research Group, H-6500 Baja, Szegedi \'ut, Kt. 766, Hungary 
         \and
   	   Baja Astronomical Observatory of University of Szeged, H-6500 Baja, Szegedi \'ut, Kt. 766, Hungary.\\
             \email{borko@bajaobs.hu} 
	\and
	   Konkoly Observatory, Research Centre for Astronomy and Earth Sciences,  H-1121 Budapest, Konkoly Thege Mikl\'os \'ut 15-17, Hungary	   
	\and
	  Department of Physics, Kavli Institute for Astrophysics and Space Research, M.I.T., Cambridge, MA 02139, USA \\
	   \email{sar@mit.edu}	  
	\and
	  Astronomical Institute, Faculty of Mathematics and Physics, Charles University, V~Hole\v{s}ovi\v{c}k\'ach 2, CZ-180~00, Praha 8, Czech Republic
	\and
           FZU - Institute of Physics of the Czech Academy of Sciences, Na Slovance 1999/2, CZ-182 21, Praha, Czech Republic 
	\and
	   Research Centre for Theoretical Physics and Astrophysics, Institute of Physics, Silesian University in Opava, Bezru\v{c}ovo n\'am. 13, CZ-746 01, Opava, Czech Republic
	\and
	   Astronomical Institute, Academy of Sciences, Fri\v{c}ova 298, 251 65, Ond\v{r}ejov, Czech Republic
	\and
	   Department of Experimental Physics, University of Szeged, H-6720, Szeged, D\'om t\'er 9, Hungary
	\and
	   NASA Goddard Space Flight Center, 8800 Greenbelt Road, Greenbelt, MD 20771, USA 
	\and
	   SETI Institute, 189 Bernardo Avenue, Suite 200, Mountain View, CA 94043, USA
       }

   \date{Received  .., 2026; accepted ... , ...}
 
  \abstract
  {}
   {Several hundred previously unknown 2+2 hierarchical doubly eclipsing quadruple stars have been identified in observations made with the space telescope TESS over the past eight years. Moreover, as of now, there have been more than a hundred close hierarchical triple star systems, identified in the same TESS datasets, with an inner eclipsing binary and a tertiary star that either eclipses the inner binary, and/or vice versa. In this work we identify and photodynamically analyze the very first 2+2-type quadruple stellar system which is the member of both groups. That is, TIC 433545934 consists of two eclipsing binaries that are revolving around each other forming a 2+2 type quadruple system in which, additionally, one binary eclipses the stars of the other binary.  One such outer eclipse was discovered with TESS, which triggered a special interest in this system.}
   {Most of the data for this study come from TESS observations, but we also obtained supplemental ground-based photometric measurements for this quadruple system. The eclipse timing variation curves extracted from the TESS and ground-based follow-up data, the photometric light curves, and the spectral energy distribution are combined in a complex photodynamical analysis to yield the stellar and orbital parameters of this dynamically interesting system.}
   {The periods of the two inner, eclipsing binaries were found to be $P_\mathrm{A}=2\fd07$ and $P_\mathrm{B}=1\fd41$, while the mutual or, outer period is $P_\mathrm{AB}=224\fd5$. The outer period alone makes this system the fifth most compact known 2+2-type quadruple. Moreover, what really makes TIC~433545934 unique is that TESS observed a triple-dipped extra eclipsing event when the stars of binary B eclipsed the two components of binary A. We identified similar extra eclipsing events in the archival, ground-based data of ASAS-SN and ATLAS, as well. We found, however, that these extra events occur only once during an outer revolution, that is, binary A does not eclipse the stars of binary B. This is in accord with our finding that the outer eccentricity is quite high, being $e_\mathrm{AB}=0.62$. Our analysis reveals that binary A consists of two quite similar, but slightly evolved late A-type stars ($q_\mathrm{A}=0.92$), while in the case of binary B, the primary star, whose mass is between the masses of the two stars of binary A, is quite dominant, with $q_\mathrm{B}=0.60$. The system was found to be substantially flat, with mutual inclination angles below $\approx2\degr$. 
 }
   {}


   \maketitle
   
   \nolinenumbers

\section{Introduction}  
\label{sec:intro}

Multiple stellar systems, or any such systems where the masses are comparable, are expected to form hierarchical configurations in order to be dynamically stable.  In the case of a triple star system, which is the simplest of the multi-stellar systems, one of the three mutual separations of the components always remains much smaller than the other two separations. In other words, a hierarchical triple stellar system can be approximated by two binary systems, revolving on two (perturbed) Keplerian orbits. The inner binary is formed by the two stars that orbit around their centre of mass with a period which is much shorter than the period of the outer `binary'. The latter is formed by the third, most distant component, and the centre of mass of the inner pair, as if the sum of the masses of the inner binary components were located at their centre of mass. Naturally, the `members' of this outer binary rotate around the centre of mass of the entire triple system. 

Regarding quadruple star systems, they come in two substantially different configurations. These are the 3+1 or, more accurately, (2+1)+1 type, and the 2+2 type configurations. Doubly eclipsing quadruple stars, that is, those systems which are formed by two eclipsing binaries (EBs), naturally belong to the second group, and can be discovered easily via a blend of two EBs in the very same observed light curve. Here a problem, which naturally occurs, is that when one observes such a blended light curve, one cannot immediately decide whether the two blended EBs are physically bound together, i.e., actually forming a 2+2-type quadruple system, or whether what is seen is simply a chance alignment without any physical connection between the two EBs.  The same question, in general, does not arise when third-body, or outer eclipses occur in an observed light curve. This is because of the characteristic, complex, and often rapidly varying shapes of the extra eclipses in some light curves, in which case one can conclude without any further investigation that what is seen is a third-body eclipse in a triply eclipsing triple stellar system.

Returning to the blend of two EBs in the same light curve, the usual method for checking whether the EBs are bound is to check the eclipse timing variations (ETV) of the EBs. If they exhibit non-linear behaviour (that is, variations in the eclipsing periods) in both systems with the same period, then one may correctly infer that the two EBs are gravitationally connected to each other.  Note that if the $P_\mathrm{out}/P_\mathrm{in}$ ratio for the two EBs is sufficiently large that the physical variations in period (occurring from the mutual gravitational perturbations) can be neglected, then the two ETVs will be dominated by the pure geometrical light-travel time effect (LTTE), i.e., the mutual Roemer-delay.  In that case, the sinusoidal ETVs of the two EBs must be strictly anticorrelated with respect to each other, and the ratio of the amplitudes of the curves must be simply the inverse of the total masses of the two binaries--analogous to the case of an SB2 binary, where the amplitude ratio of the two radial velocity (RV) curves directly yields the inverse of the mass ratio of the binary.

In the case of TIC\,433545934, as will be shown in this work, both binaries' ETVs exhibit the same periodicity, making it clear that the two EBs form a short outer period quadruple system. On the other hand, however, what makes this quadruple unique, and also provides an immediate additional proof of the bound nature of these two EBs, is our identification of a three-dipped extra eclipsing event in the TESS sector 38 light curve between BJD (barycentric Julian Date) 2459356.9 and 2459358.2. Later we found additional extra eclipses, that is, mutual eclipses of the two EBs using archival and recent ground-based survey data. Therefore, TIC~433545934 is the very first such known 2+2 doubly eclipsing quadruple stellar system that shows also mutual eclipses of the two EBs.

We also note the related case of KIC\,5255552.  This system was discovered with the \textit{Kepler} spacecraft \citep{borucki10} as a long-period EB with $P_\mathrm{ecl}=32\fd45$-d, which also exhibits third-body eclipses.  From these, as well as from an ETV-analysis, \citet{borkovitsetal15} first found an eccentric outer orbit with a period $P_\mathrm{out}=862$\,d.  Due to the quite peculiar, four-dipped outer orbit eclipse, however, it was already mentioned in \citet{getleyetal20} that the `third' body itself might be another binary, thereby making this a 2+2-type outer eclipsing quadruple system.   Later, \citet{orosz23} found a reliable four-body photodynamical solution for the system, and explained the outer eclipses with the presence of another long-period ($P_\mathrm{B}=33\fd7$), {\sc non-eclipsing} binary. In this manner, KIC\,5255552 would be the first known 2+2-type quadruple star system which exhibits outer eclipses. On the other hand, however, this latter system is clearly not a doubly eclipsing 2+2 quadruple. Moreover, our current finding, in the case of TIC\,433545934, is also much more robust, because the observations make it clearly evident that this system is actually an outer-eclipsing 2+2-type doubly eclipsing quadruple.

This paper is structured as follows. In Sect. 2 we describe the discovery and further follow-up observations of TIC\,433545934. Then, the investigation of the available ground-based survey observations are discussed in Sect. 3. In Sect. 4 we discuss the study of the outer period that is mainly based on the ETVs formed from the TESS light curve, as well as from the ground-based follow-up observations. The details and results of the comprehensive photodynamical analysis are given in Sects. 5 and 6, while the paper concludes with some further discussions and summary in Sect. 7.
   
\section{Discovery and further observations of TIC 433545934}  
\label{sec:observations}

\begin{figure}[ht]
   \centering
\includegraphics[width=0.47\textwidth]{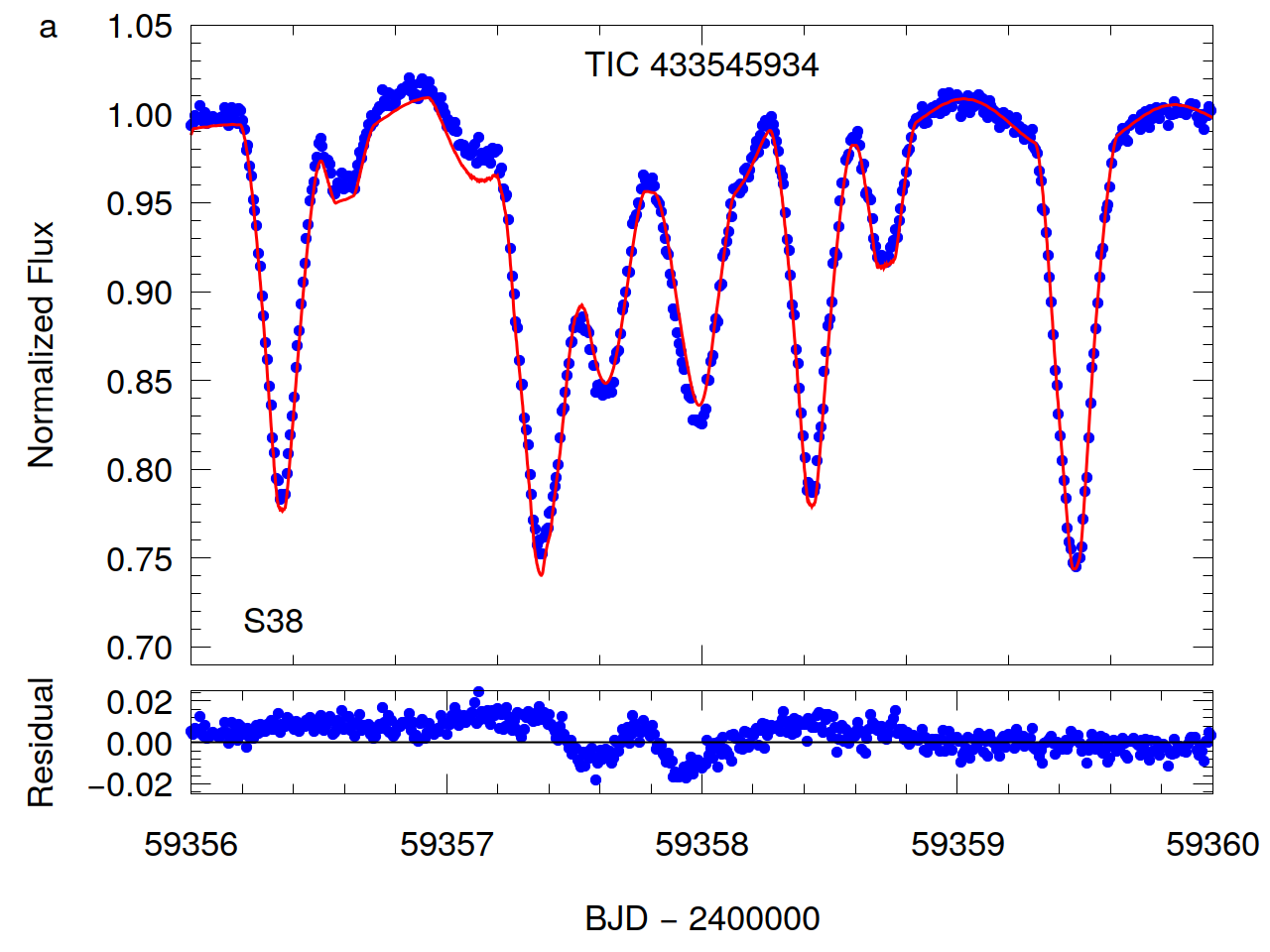}
\includegraphics[width=0.47\textwidth]{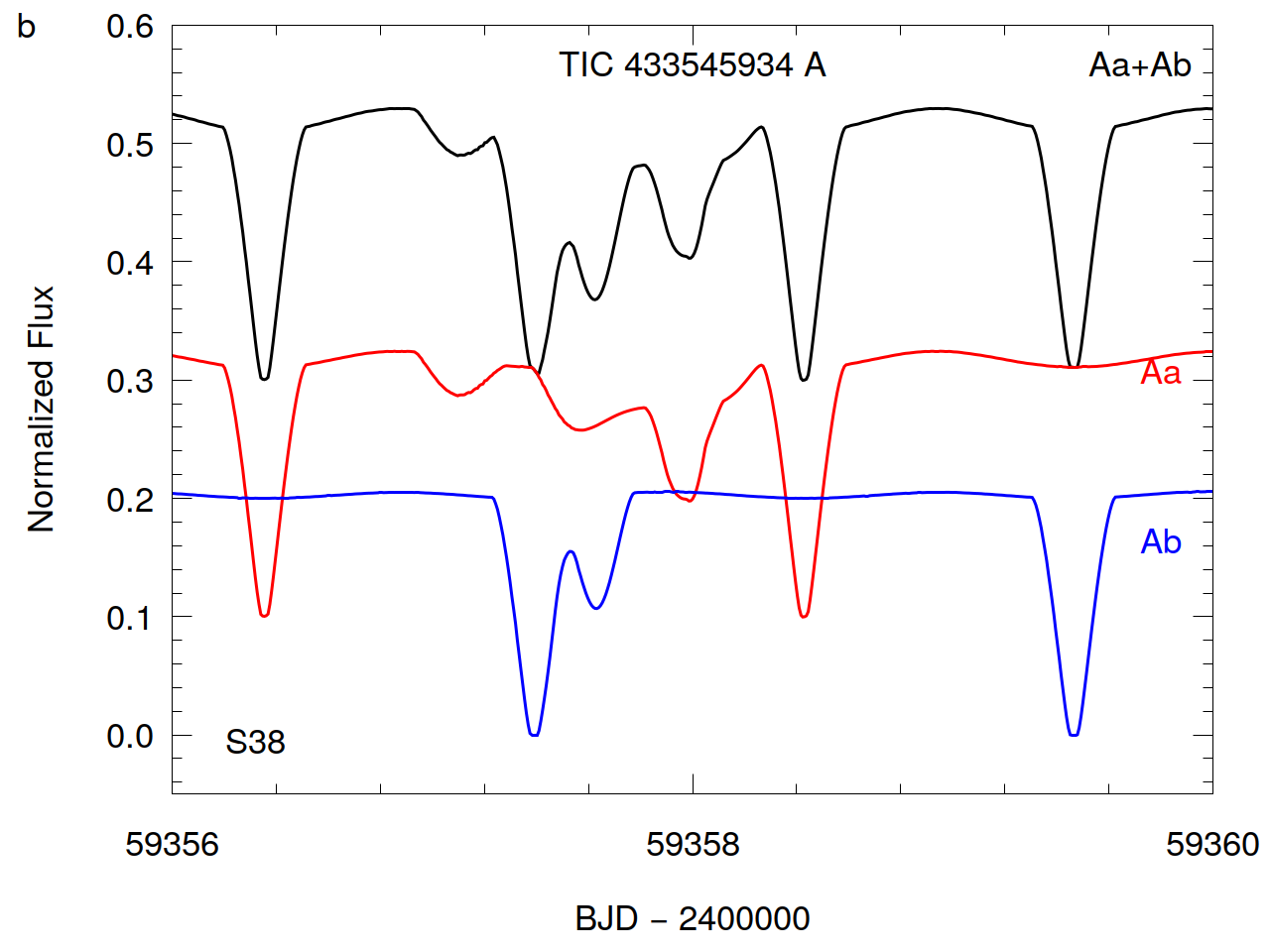}
   \caption{Panel a): The extra eclipse section of the blended light curves of binaries TIC 433545934 A and B. Blue dots represent the TESS observations, while the continuous red curve represents the best-fit photodynamical solution. The residual light curve of the fit can be seen in the lower sub-panel. Panel b): The model flux contribution of binary A (that is, stars Aa+Ab in black), as well as the primary (Aa in red) and secondary (Ab in blue), during the extra eclipse. The three extra dips in flux caused by the outer eclipses of the components of binary B can be readily seen. The vertical axis shows the normalized flux contribution of the corresponding objects. Note that the flux contribution of Ab (blue) during the flat bottom of its occultations (that is, during the secondary eclipses of binary A), is exactly zero, since during these phases star Ab is completely occulted by star Aa.}
   \label{fig:433545934Aoutereclipse}
\end{figure}

TIC\,433545934 (ASASSN-V J124203.23-644513.2; 2MASS J12420321-6445132, see Table~\ref{tbl:mags} for the main catalog parameters) was observed with the TESS space telescope \citep{ricker15} during six different sectors in the full frame image (FFI) mode. In the case of the Sector 38 observations, the observation cadence time was 600 sec, while for Sectors 64, 65, 99-101, the cadence was 200-sec.  Both the doubly eclipsing nature, and likely third-body eclipse in the Sector 38 data, were first reported in \citet{zascheetal22}, and can be seen nicely in the section of the light curve that is plotted in Fig.~\ref{fig:433545934Aoutereclipse}. After that time, we added this target to the potentially interesting systems that we study.  Only after the three most recent sectors of TESS observations, however, did we feel that we had gathered sufficient material for a detailed analysis and evaluation of the system.

We processed the Year 3 and 5 TESS FFI light curves, with the {\sc FITSH} package \citep{pal12}, similar to what we did in our prior work. On the other hand, the new Year 7 TESS observations (that is, the S99-101 FFI data) were processed with the publicly available software {\sc Lightkurve} \citep{2018ascl.soft12013L}, but the eclipse depths in this new data set were scaled to the older ones.

We also carried out some additional, ground-based photometric observations. These are as follows: \\
-The telescope FRAM is an Orion ODK 300/2040\,mm, equipped with an MII G4-16000 CCD camera. The location of the FRAM telescope itself is at the peak of Los Leones, at the Pierre Auger Observatory, near the town of Malarg\"ue, 
Argentina \citep{aabetal21}. \\
-The 1.54 m Danish telescope (DK154) is a Ritchey-Chretien type, built in 1979, located at La Silla observatory in Chile. It is equipped with a DFOSC CCD camera, and the data were obtained in the standard R-band filter.

Since no further extra eclipses were observed and, moreover, we were unable to find or obtain any RVs for the constituent stars, we followed our two usual methods for finding any reasonable outer orbital periods. These two methods depend on (i) the use of archival data and (ii) the determination of the mid-minima times of the regular eclipses.  After that the analysis of the eclipse timing variations (ETVs) of both binaries.

\begin{table}[ht]
\hspace{-20px}
\centering
\caption{Main properties of TIC 433545934 from different catalogs}
\begin{tabular}{lc}
\hline
\hline
Parameter			       &   \\  
\hline
RA J2000 			       &  12:42:03.18 \\
Dec J2000 		               & $-$64:45:13.48 \\
$T$\tablefootmark{a}                   & $12.40 \pm 0.01$ \\
G\tablefootmark{b} 		       & $13.06 \pm 0.00$ \\ 
G$_{\rm BP}$\tablefootmark{b} 	       & $13.53 \pm 0.01$ \\
G$_{\rm RP}$\tablefootmark{b} 	       & $12.31 \pm 0.01$ \\
B\tablefootmark{c}                     & $14.18 \pm 0.17$ \\
V\tablefootmark{c}                     & $13.39 \pm 0.16$ \\  
J\tablefootmark{d}		       & $11.36 \pm 0.04$ \\
H\tablefootmark{d} 		       & $11.06 \pm 0.06$ \\
K\tablefootmark{d} 		       & $10.72 \pm NA$  \\
W1\tablefootmark{e} 		       & $10.35 \pm 0.02$  \\
W2\tablefootmark{e} 		       & $10.38 \pm 0.02$ \\
W3\tablefootmark{e} 	               & $10.99 \pm 0.10$  \\
$T_{\rm eff}$ [K]\tablefootmark{b}     & $7861 \pm 253$ \\  
Distances [pc]\tablefootmark{f}	       & $4105\pm 1567$ \\ 
                                       & $2392\pm 699$ \\
$E(B-V)$\tablefootmark{a}              & NA \\ 
$\mu_\alpha$ [mas/yr]\tablefootmark{b} & $-9.63 \pm 0.18$ \\   
$\mu_\delta$ [mas/yr]\tablefootmark{b} & $-1.49 \pm 0.22$ \\  
RUWE\tablefootmark{b,g}                & 14.85  \\
astr\_ex\_noise [mas]\tablefootmark{b,h}&  2.18 \\
astr\_ex\_noise\_sig\tablefootmark{b,h}&  4125 \\
\hline
\label{tbl:mags}
\end{tabular} 

\small
\textbf{Notes.} General: ``NA" indicate that the value is not available. (a) TESS Input Catalog (TIC v8.2) \citep{TIC8}. (b) Gaia EDR3 \citep{GaiaEDR3}; the uncertainty in $T_{\rm eff}$ listed here is 1.5 times the geometric mean of the upper and lower error bars of \verb|teff_gspphot|. Magnitude uncertainties listed as 0.00 are $\lesssim 0.005$. (c) AAVSO Photometric All Sky Survey (APASS) DR9, \citep{APASS}, \url{http://vizier.u-strasbg.fr/viz-bin/VizieR?-source=II/336/apass9}. (d) 2MASS catalog \citep{2MASS}.  (e) WISE point source catalog \citep{WISE}. (f) \citet{bailer-jonesetal21}, geometric and photogeometric distances (upper and lower rows, respectively). The uncertainties are calculated as the geometric means of the upper and lower error bars. (g) The Gaia renormalized unit weight error (RUWE) is the square root of the normalized $\chi^2$ of the astrometric fit to the along-scan observations. Values in excess of about unity are sometimes taken to be a sign of stellar multiplicity. (h) Abbreviations for \verb|astrometric_excess_noise| and \verb|astrometric_excess_noise_sig| (\citealt{lindegren21}; \url{https://gea.esac.esa.int/archive/documentation/GDR2/Gaia_archive/chap_datamodel/sec_dm_main_tables/ssec_dm_gaia_source.html}); these are a measure of ``the disagreement, expressed as an angle, between the observations of a source and the best-fitting standard astrometric model.'' Values of \verb|astrometric_excess_noise_sig| $\gtrsim 2$ are considered significant.
\end{table}

\section{Investigation of the archival data}
\label{sec:archive}

In order to obtain a first estimate of the outer period in TIC 433545934, we made use of the achival photometric data from the Asteroid Terrestrial-impact Last Alert System (ATLAS; \citealt{tonry18}) and from the All-Sky Automated Survey for SuperNovae (ASAS-SN; \citealt{shappee14}; \citealt{kochanek17}).  The idea is to search for repeated occurrences of the outer eclipses in the archival data.  To this end, we downloaded approximately 5000 photometric values spanning the past decade from ASAS-SN\footnote{\url{https://asas-sn.osu.edu/}}, and about 2000 values from the past 4.4 years from ATLAS\footnote{\url{https://fallingstar-data.com/forcedphot/queue/}}.

The binary periods of binary A and binary B are immediately apparent from Box Least Squares (BLS) transforms \citep{kovacs02} of both data sets: $P_A = 2.07250 \pm 0.00001$ d, and $P_B = 1.41224 \pm 0.00002$ d.  In order to be able to detect possible third body eclipses, which, in this case, are no deeper than the EB eclipses, but have much lower duty cycle, we must remove the EB lightcurve profiles from the lightcurve.  We first co-add the data from ATLAS and ASAS-SN after median normalizing the magnitudes obtained in the two different filter bands of each survey. Once this is done we simultaneously fit for the first 50 harmonics of each binary orbital period.  In all there are 201 coefficients to fit for (a sine and cosine term for 50 harmonics times 2 orbital periods, plus a constant term).  With these linear coefficients in hand, the functional forms representing the two orbital lightcurves are subtracted from the lightcurve.  Finally the cleaned data set is subjected to a BLS transform \citep{kovacs02} to search for the eclipses of the outer period.

\begin{figure}[ht]
   \centering
\includegraphics[width=0.475\textwidth]{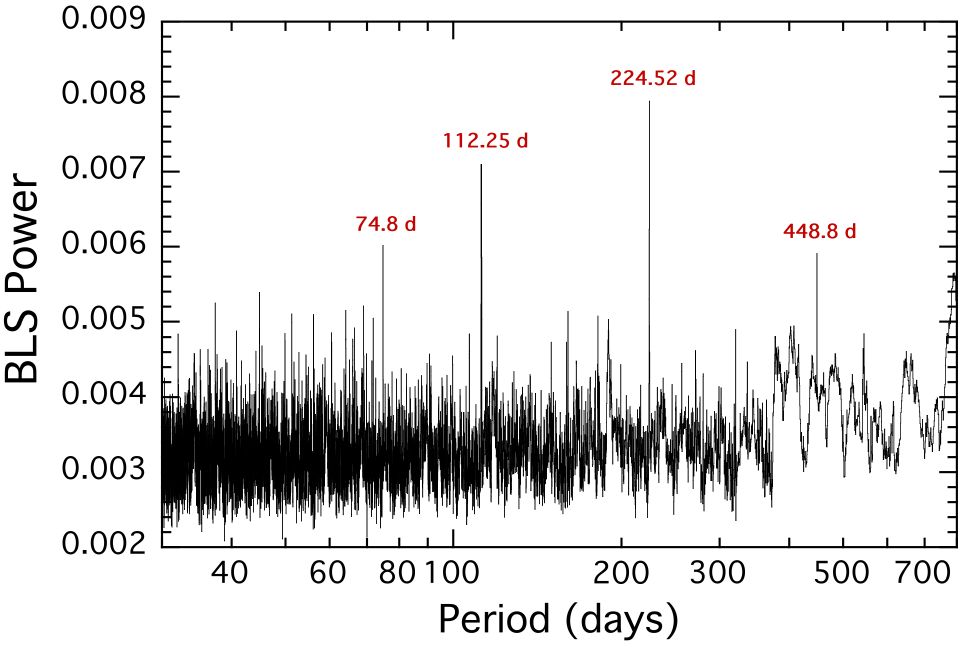} \hglue0.4cm
\includegraphics[width=0.47\textwidth]{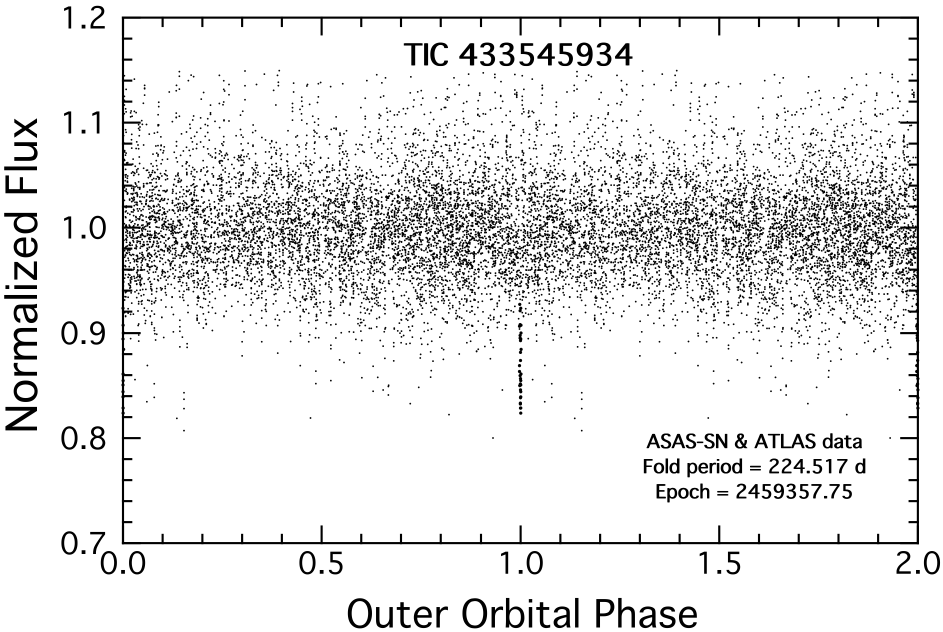}
   \caption{TIC 433545934 outer orbit eclipses in the archival data.  Top panel a):  BLS transform of the archival ATLAS and ASAS-SN data -- some 7000 points in total. The largest peak yields a period for the outer orbit of 224.52 days; the other large peaks are harmonics. Panel b): Pointwise fold of the archival data about a period of 224.52 days, and phased to the one outer eclipse observed with TESS. The low points from the outer eclipse are highlighted with slightly darker points.}
   \label{fig:archival}
\end{figure}

The results of the BLS transform of the combined, normalized, and cleaned ATLAS and ASAS-SN data sets are shown in the upper panel of Fig.~\ref{fig:archival}.  There are four significant periods between 30 d and 800 d, the highest of which is at $224.52 \pm 0.03$ d.  The other peaks are twice, 1/2, and 1/3 this value, i.e., harmonics.  The bottom panel in Fig.~\ref{fig:archival} shows a pointwise fold of the data about this period.  Phase zero in this plot corresponds to the time of the third body eclipse seen in the TESS data.  As is evident, there is a significant quite narrow dip ($\sim$$\pm 0.5$ d; $\pm 0.002$ in phase) containing some 28 low flux values (highlighted with darker points).  On average, other phase bins this narrow have only 1 or 2 low points. 

From this study of the achival data we conclude that (i) we have indeed detected the outer eclipses in this quadruple, and (ii) the very likely outer period is 224.52 d, but (iii) we cannot absolutely rule out a period of half that value from the archival data alone. 

\section{Period study}  
\label{sec:periodstudy}

As was mentioned above, another way to obtain the outer orbital period is via a determination of the times of eclipse minima, especially from the well-covered, accurate, and dense TESS data, followed by an investigation of the corresponding ETV data.  In order to obtain mid-eclipse times for both inner binaries, it was necessary to obtain disentangled light curves for both binaries. This disentanglement was done in the identical way as was described previously in \citet{zascheetal23}.  Such a disentanglement before the eclipse-time calculations was exceptionally important in the current situation since, due to the short periods of both inner binaries, there were several overlapping eclipses. Therefore, it was critical to obtain such light curves for both binaries which did not contain the signal of the other binary. Moreover, naturally, the $\sim$$2.5$-day-long portion of the only TESS-observed outer eclipse was eliminated from our ETV analysis.

\begin{figure*}[ht]
   \centering
\includegraphics[width=0.47\textwidth]{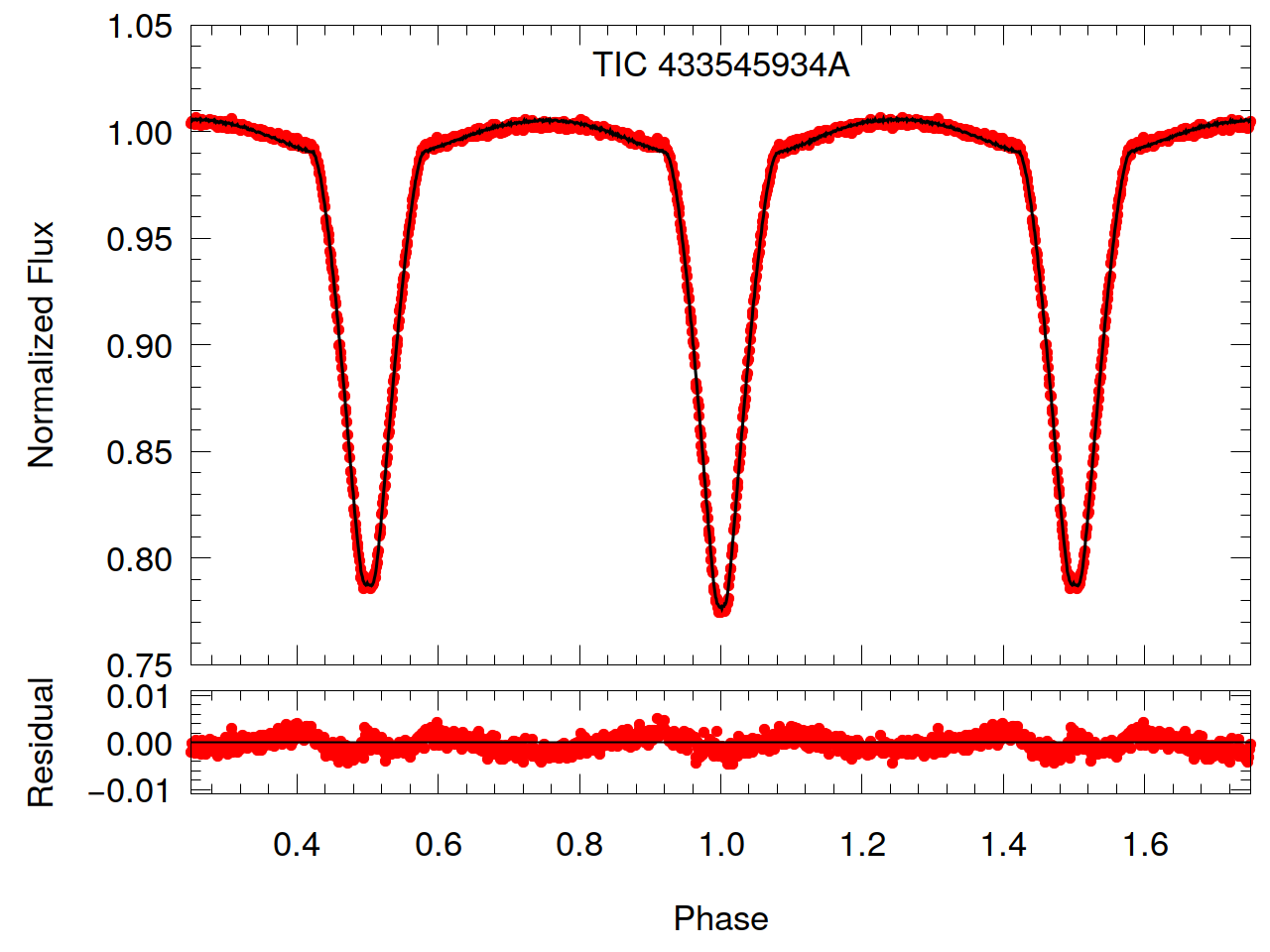}\includegraphics[width=0.47\textwidth]{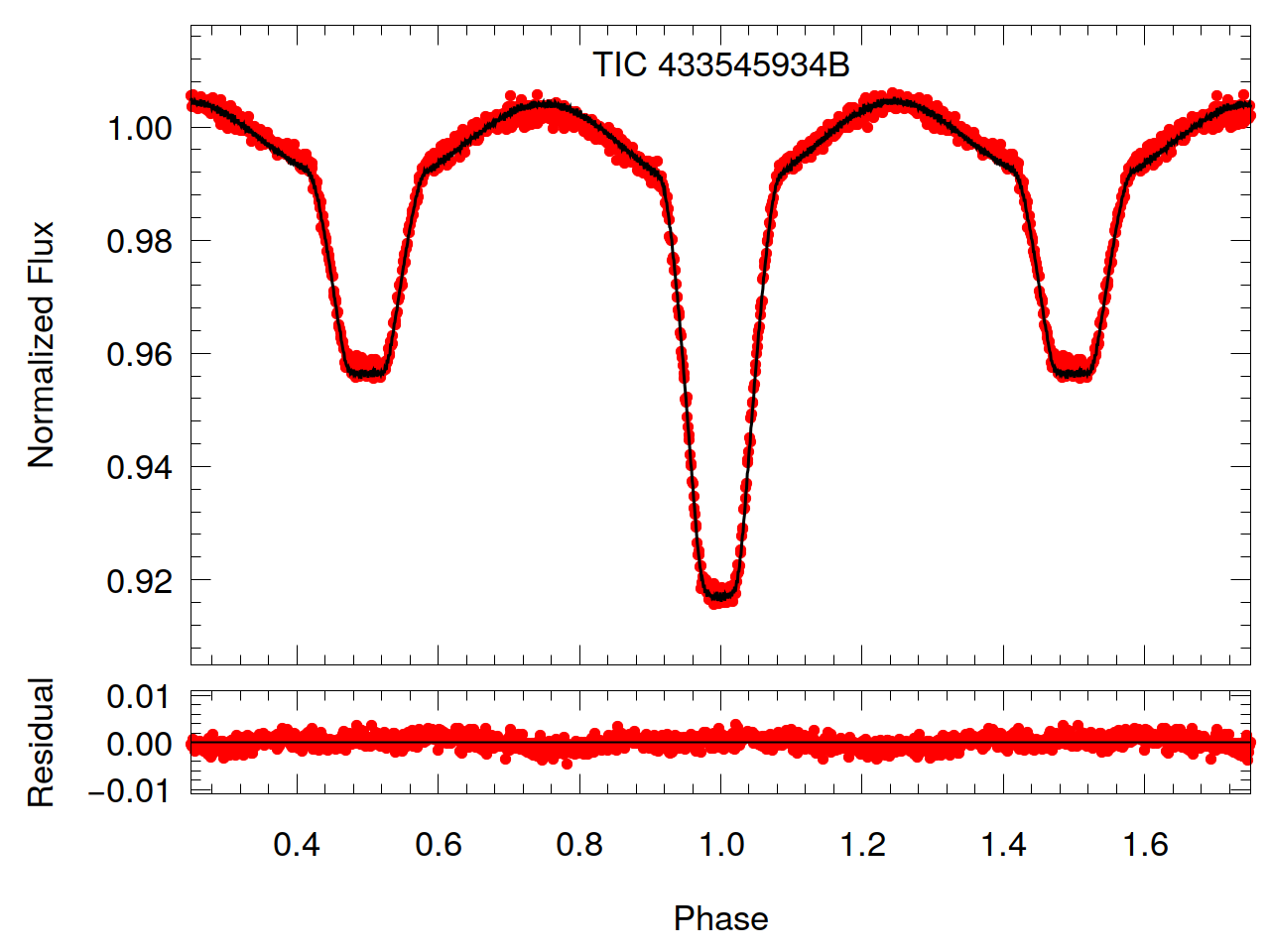}
   \caption{The disentangled, folded, and binned light curves of TICs 433545934A and B (left and right panels, respectively). The red dots in the upper panels represent the phase-binned TESS observations. The black lines are the disentangled, folded, and binned light curves of the entire best-fit photodynamical model. (Therefore, these are not direct fits of the folded TESS light curves.) Similarly, in the bottom panels, the disentangled, folded, and binned residual light curves of the entire photodynamical model can be seen.}
   \label{fig:433545934ABlcfold}
\end{figure*}

\begin{figure}
\centering
     \includegraphics[width=\hsize]{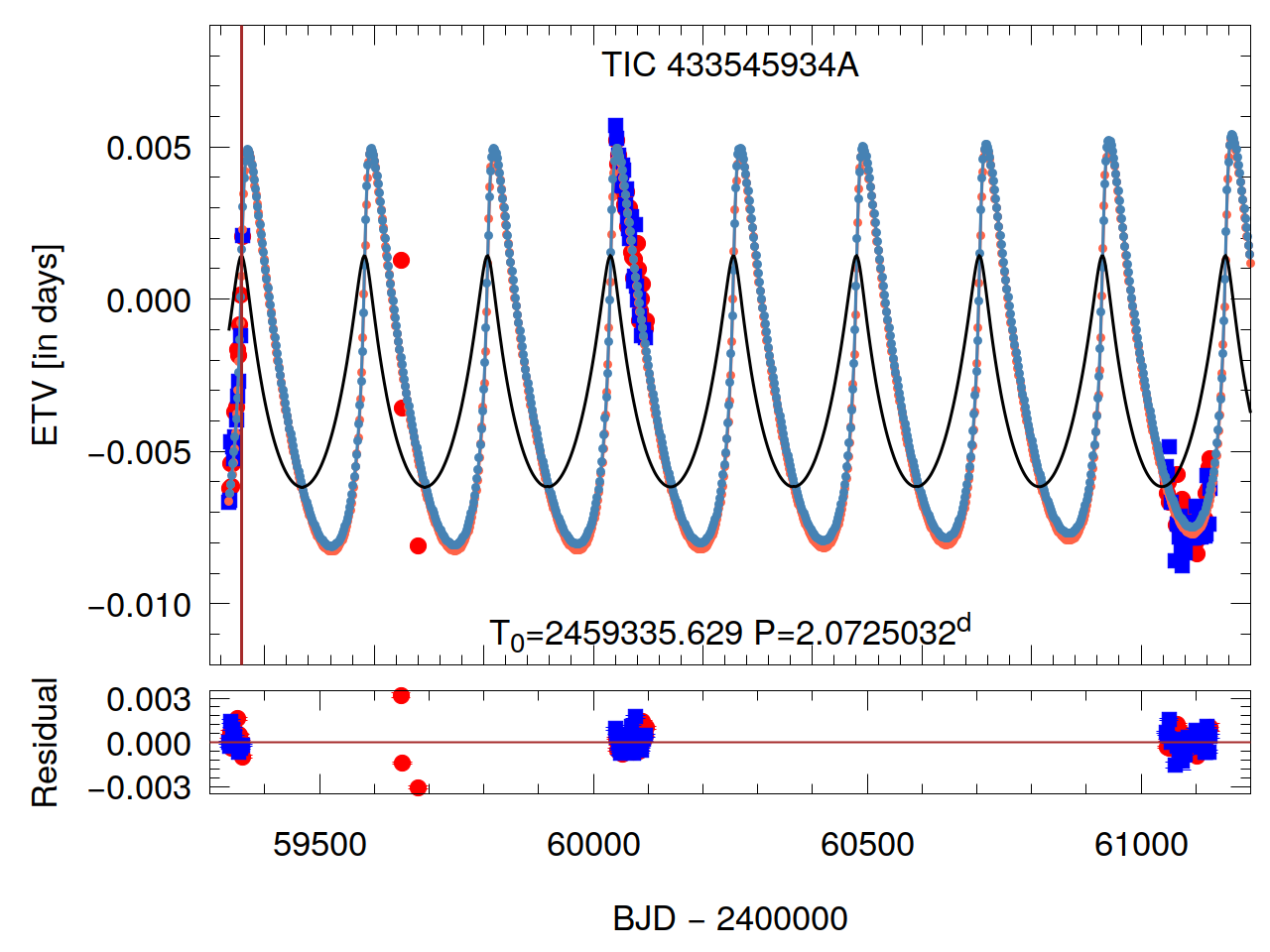}
     \includegraphics[width=\hsize]{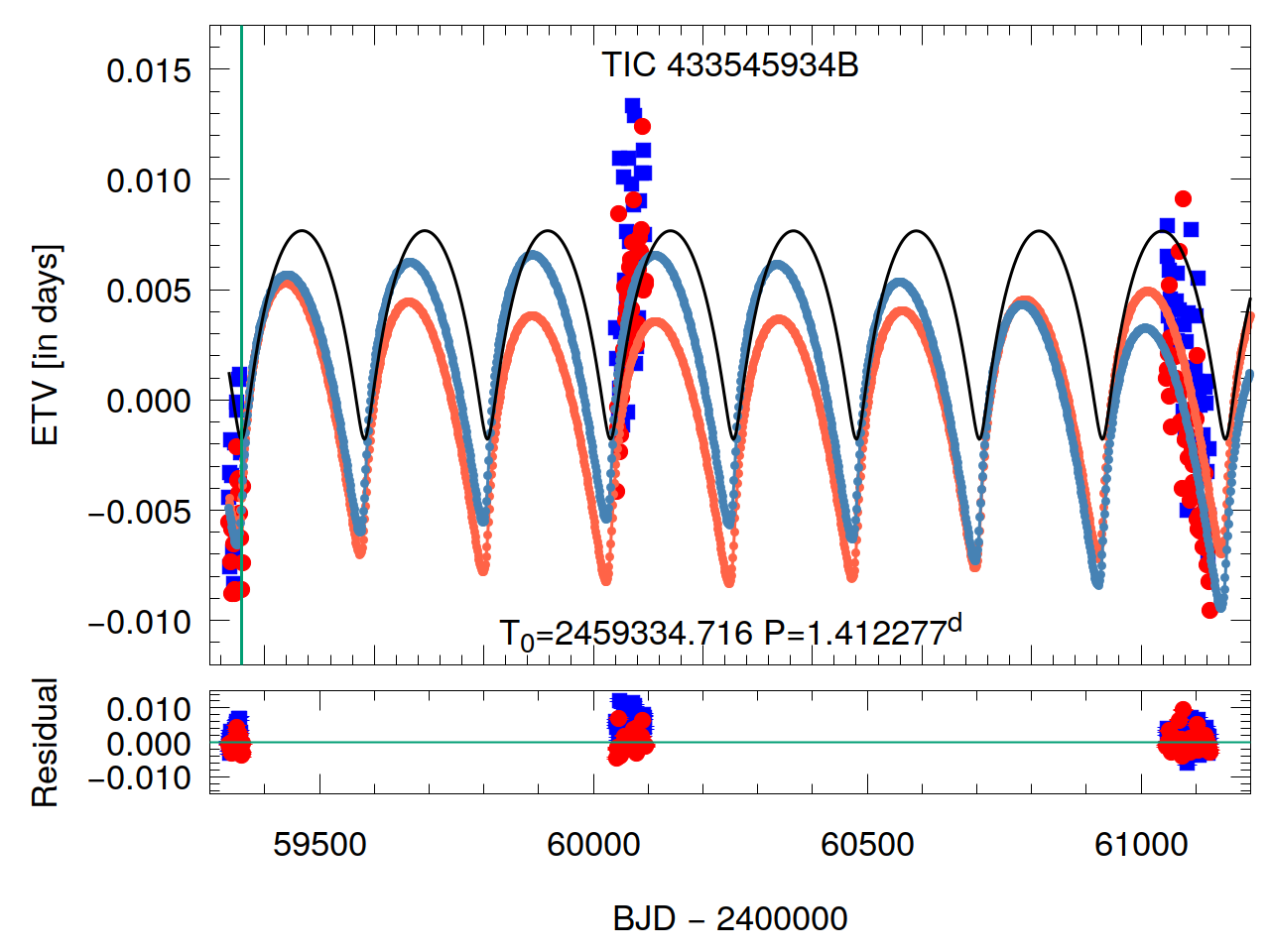}
     \caption{Primary and secondary ETV curves (heavy red and blue circles, respectively) formed from the TESS observations of TICs 433545934A and B (top and bottom, respectively). The small orange and cyan points, connected by thin lines, represent the best-fit photodynamical solution.  The black curves show the pure LTTE contribution to both ETVs. As one can see, the black curves are anticorrelated in the case of the ETVs from the two EBs, and their amplitude ratio gives the reciprocal of the outer mass ratio. Vertical lines mark the time of the only observed outer eclipse, which is located nicely at the upper and lower extrema of the two LTTE contributions.}
\label{fig:433545934ABETV}
\end{figure}

For illustration, we show in Fig.~\ref{fig:433545934ABlcfold} the phase curves obtained from the disentangled EB light curves after they were folded with the disentangling period and binned into 1000 equally spaced phase elements. (Note, however, these folded light curves were used only to form templates for all four types of different eclipses -- primary and secondary minima in both binaries, -- but the individual times of eclipse minima were determined from the disentangled time series.) These individual mid-eclipse times are tabulated in Tabs.~\ref{Tab:TIC_433545934A_ToM} and \ref{Tab:TIC_433545934B_ToM}. From these times, ETV curves were formed for both binaries (see in Fig.~\ref{fig:433545934ABETV}). The ETV curves had different slopes during the different years the source was observed, providing some newer independent proof of the short-timescale period variations. On the other hand, they did not reflect clearly anticorrelated behaviour; but, we emphasize again, that feature is expected only in the case of purely LTTE dominated ETVs.

In the case of the determination of additional mid-eclipse times from the ground-based follow-up data, we applied a somewhat different method. For these observations, which yielded only short segments of data, we had no chance of doing a proper disentanglement of the signals of the two EBs. Therefore, we calculated further times of minima only from those eclipse observations where, according to the pre-calculated photometric phase values, no overlapping of the eclipses of the two binaries was expected to occur. Moreover, we did not use those ground-based follow-up observations for mid-eclipse time determination, when simultaneous TESS-observations were also available. In this manner we were able to determine three additional mid-minima times for binary A --- these are also listed in Tab.~\ref{Tab:TIC_433545934A_ToM}. In the case binary B, with the more shallow eclipse depths, we were unable to determine any additional mid-minima times from the ground-based observations.

We then carried out a preliminary analytic ETV analysis for a coplanar orbit with the code described in \citet{borkovitsetal15} for the dominant binary A. This revealed the following main parameters: $P_\mathrm{AB}=225.02\pm0.25$\,days, $e_\mathrm{out}=0.57\pm0.01$, $\omega_\mathrm{out}=261\degr\pm4\degr$, $M_\mathrm{A}=4.5\pm0.2\,\mathcal{M}_\sun$, $M_\mathrm{B}=4.2\pm0.3\,\mathcal{M}_\sun$, while the ratio of the amplitudes of the (quadruple) dynamical and the LTTE terms was found to be $\mathcal{A}_\mathrm{dyn}/\mathcal{A}_\mathrm{LTTE}=1.11$. At this point we emphasize that even with our analytic ETV fit we could already clearly break the ambiguity (see in Sect.~\ref{sec:archive}) between the outer periods of $P_\mathrm{out}\sim224\fd52$ and half this value of $P_\mathrm{out}\sim112\fd26$. In particular, for this latter value of 112 d, our analysis did not give any physically reasonable solution(s). Therefore, we carried out complex photodynamical analysis in which the inner parameters of the several MCMC runs were set to be close to these values.

\section{Photodynamical model}
\label{sec:photodynamics}

Similar to what was done in our former works, both on triply eclipsing triple stars and doubly eclipsing, 2+2 type quadruple systems, TIC\,433545934 was subjected to a detailed photodynamical analysis with the use of our own developed software package {\sc Lightcurvefactory}. We have described the details of such a photodynamical analysis in numerous former papers \citep[see, e.g.][]{borkovitsetal18,borkovitsetal19a,borkovitsetal19b,borkovitsetal20a,borkovitsetal20b,borkovitsetal21,mitnyanetal20}. In these former papers we described in detail all the parameters we adjust or constrain during each MCMC run. We simply refer to these works to save us from repeating all this information here, and mention only that, in the current situation, we adjusted and constrained 22 and 15 variables, respectively.  The adjusted variables include the primary's (i.e., star $Aa$) mass and the other mass ratios; three, four and six orbital elements or other parameters of the three (two inner and one outer) orbits, respectively; passband-dependent extra fluxes; and age, metallicity and interstellar extinction of the system. The constrained parameters include the stellar radii and effective temperatures from the built-in \texttt{PARSEC} evolutionary tracks, as well as the remaining orbital parameters, and finally the photometric distance of the quadruple star, and the passband-dependent limb darkening coefficients of the four stars. All the parameters were adjusted or constrained in the same manner as described in the papers listed above. In the case of most of the adjusted parameters, uniform priors were used, with the exception of such relative quantities as the mass ratios, in which case, logarithmic priors were applied. We point specifically to the two schematic flow-charts given as Fig.\,5 of \citet{borkovitsetal20a} and Fig.\,3 of \citet{borkovitsetal25}.
Here we note that our MCMC approach is an implementation of the generic Metropolis-Hastings algorithm \citep[see, e.g., ][]{ford05}. During the fitting process, we initiated several dozens of different chains. All contained a hundred thousand trial steps, and each chain stopped when it reached ten thousand accepted steps. In order to neglect the burn-in phase of each chain we considered only those trial steps for the final statistics where the global $\chi^2$ value of the given trial run did not exceed 120\% of the lowest $\chi^2$ value of all the runs (that is, the $\chi^2$ of the so-called `best-fit' solution).

\begin{table}
\centering
\caption{Definitions of the system parameters in Table~\ref{tab:syntheticfit_TIC433545934AB}}
\label{tbl:definitions}
\small
\begin{tabular}{lc}
\hline
\hline
Parameter\tablefootmark{a} & Definition   \\
\hline
$t_0$ & Epoch time for osculating elements    \\
$P$ & Orbital period  \\ 
$a$ & Orbital semimajor axis  \\
$e$ & Orbital eccentricity \\
$\omega$ & Argument of periastron (of secondary) \\
$i$ & Orbital inclination angle \\
$\mathcal{T}_0^\mathrm{inf}$ & Time of conjunction of the secondary \\
$\tau$ & Time of periastron passage  \\
$\Omega$ & Longitude of the node relative to \\
& the node of the inner orbit \\
$i_{\rm mut}$ & Mutual inclination angle\tablefootmark{b} \\
$q$ & Mass ratio (secondary/primary)  \\ 
$K_\mathrm{pri}$ & ``K'' velocity amplitude of primary \\
$K_\mathrm{sec}$ & ``K'' velocity amplitude of secondary \\
$R/a$ & Stellar radius divided by semimajor axis \\
$T_{\rm eff}/T_{\rm eff,Aa}$ & Temperature relative to EB primary \\
fractional flux  & Stellar contribution in the given band \\
$M$ & Stellar mass  \\
$R$ & Stellar radius   \\
$T_\mathrm{eff}$ & Stellar effective temperature  \\ 
$L_\mathrm{bol}$ & Stellar bolometric luminosity  \\
$M_\mathrm{bol}$ & Stellar absolute bolometric magnitude \\
$M_V$ & Stellar absolute visual magnitude \\
$\log g$ & log surface gravity (cgs units) \\
$[M/H]$ & log metallicity abundance to H, by mass \\
$E(B-V)$ & Color excess in B-V bands  \\
extra light, $\ell_x$  & Contaminating flux in the given band   \\
$(M_V)_\mathrm{tot}$ & System absolute visual magnitude   \\ 
distance & Distance to the source  \\
\hline   
\end{tabular}
\tablefoot{(a) The units for the parameters are given in Table~\ref{tab:syntheticfit_TIC433545934AB}. (b) More explicitly, this is the angle between the orbital planes of the two inner binaries and, moreover, between the given binary and the outer orbit.}
\end{table}


Regarding the mass determination of the currently investigated quadruple system, we should make some additional comments. In the absence of RV data, information about the absolute values of stellar masses, temperatures, and radii was augmented by the use of the composite SED of the four stars, through \texttt{PARSEC} isochrones \citep{PARSEC} as proxies. The details of such an approach, and, moreover, its natural limitations were described in \citet{borkovitsetal22}, as well as in \citet{rappaportetal24}.

We use observed passband-magnitudes from the far ultra-violet to the near infrared passbands. In the case of the current system we had an additional problem. Using the usual WISE magnitudes, we were unable to obtain reasonable SED fits.  We found that the WISE $W_1$ and $W_2$ fluxes were slightly higher than expected, that is, for any reasonable fits, our solutions produced slightly fainter system fluxes in the near-infrared region than the observed fluxes. This fact might be explained either by the presence of nearby red stars\footnote{There are three neighbor stars at 3.7, 5.3, and 6.3 arc seconds that are fainter than TIC\,433545934 by $\Delta G$ = 3.6, 4.6, and 3.9, and with $Bp-Rp$ = 2.4, 2.2 and 2.5, respectively, compared to a value of 1.2 for TIC\,433545934. Together these are equivalent to $\sim$25\% of the flux in the $W_1$ and $W_2$ bands of TIC\,433545934, but much of this may have been removed in the processing of the WISE fluxes.  However, we have no way of assessing this.}  whose fluxes leak into the photometric aperture of TIC\,433545934, or by some infrared excess in the system itself. Therefore, we decided to ignore the $W_3$ magnitude, and we down-weighted the $W_1$ and $W_2$ magnitudes by $\sim$$50\%$.  Even with this choice, however, the total SED fit was weaker than what we usually found in our previous works (that is, the value of $\chi_\mathrm{SED}^2$ was higher than in the former works by some $\sim$$500\%$).

In conclusion we used the following observational material for our analysis: 
\begin{itemize}
\item[(i)] We binned the TESS lightcurves to 1800 sec cadence times. In our previous work, we followed a similar process, mainly for fitting  TESS light curves where the cadence times were different. In the current situation no 1800-sec cadence-time FFIs are available (as TESS did not observe this target during its main mission, where the cadence times of the FFIs were only 1800 sec). Despite this, in order to save computational time, we binned both the 600-sec and 200-sec cadence time FFIs to 1800-sec, and used this 1800-sec binned TESS data as the light curve to be fitted. This means that, in the case of the 600-sec or 200-sec cadence observations, we averaged three or nine consecutive measurements, respectively, and used these averaged values with the corresponding mid-times of these three or nine observations. In this regard, we note that the eclipses are quite wide in phase in both EBs, that is, they are 5-7 hour-long, as one can see in Fig.~\ref{fig:433545934ABlcfold}.  Therefore, even in the case of an 1800-sec-binning, each individual eclipse contains 10-14 1800-sec-binned (i.e., half-hour-binned) observational points, and all the individual eclipses are well-covered for the analysis even with the use of this coarser binning. We used also a second light curve that was made from a compilation of all the ground-based follow-up observations.  Since the original exposure times were different from night-to-night, and even for the very same instrument under different weather conditions, we binned these data to 900-sec, and used this as a second dataset. Since most of the ground-based photometry was carried out in the Cousins $R_C$ band, we assigned this filter to this second light curve. Note, however, that this light curve was taken into account in the fitting only with a weight that was ten times smaller than that of the TESS light curve. In such a manner, the light curve fitting part of our photodynamical modelling was based almost exclusively on the much more accurate TESS data, and the ground-based observations in this regard served mostly as a check by penalizing only the worst fit ground-based data in between the few TESS-observed data sections.
\item[(ii)]  As mentioned in Sect.~\ref{sec:periodstudy}, we formed mid-eclipse times from the original TESS and ground-based datasets. These times of minima, again, are listed in Tables~\ref{Tab:TIC_433545934A_ToM} and \ref{Tab:TIC_433545934B_ToM}. The shallow and flat secondary eclipses in the folded light curve of binary B (see the right panel of Fig.~\ref{fig:433545934ABlcfold}) sometimes produced ETVs with unacceptably large scatter. (This might be the consequence of an insufficient light curve disantenglement.) Therefore, we dropped out some evident outlier secondary ETV points from the ETVs of binary B. (Note that these bad ETV points are not listed in the corresponding table.)
\item[(iii)] Finally, simultaneously with the light curve and ETV curve fitting, SED fitting was also applied. The observed catalog magnitudes used for this fitting are listed in Table~\ref{tbl:mags}. Some details and caveats of the SED fitting have also been discussed in our previous works.
\end{itemize}

\section{System parameters}
\label{sec:results}

\begin{table*}
 \centering
\caption{Orbital and astrophysical parameters of TICs\,433545934 A and 433545934 B from the joint photodynamical lightcurve, ETV, SED and \texttt{PARSEC} isochrone solution. }
 \label{tab:syntheticfit_TIC433545934AB}
\scalebox{0.91}{\begin{tabular}{@{}lllll}
\hline
\hline
\multicolumn{5}{c}{Orbital elements} \\
\hline
   & \multicolumn{3}{c}{subsystem}  \\
   \cline{2-3}
   & A & B & A--B &  \\
  \hline
  $t_0$ [BJD - 2400000]& \multicolumn{3}{c}{$59333.8$}  \\
  $P$ [days] & $2.072891_{-0.000034}^{+0.000034}$ & $1.412719_{-0.000043}^{+0.000043}$ & $224.84_{-0.30}^{+0.30}$ & \\
  $a$ [R$_\odot$] & $11.29_{-0.13}^{+0.15}$ & $8.133_{-0.082}^{+0.104}$ & $312.5_{-3.3}^{+4.1}$ &  \\
  $e$ & $0.00059_{-0.00012}^{+0.00013}$ & $0.0028_{-0.0010}^{+0.0010}$ & $0.6228_{-0.0063}^{+0.0062}$ & \\
  $\omega$ [deg]& $294_{-16}^{+14}$ & $252_{-39}^{+35}$ & $273.5_{-3.8}^{+3.8}$ & \\ 
  $i$ [deg] & $87.58_{-0.42}^{+0.82}$ & $88.68_{-0.41}^{+0.61}$ & $88.92_{-0.05}^{+0.05}$ & \\
  $\mathcal{T}_0^\mathrm{inf}$ [BJD - 2400000]& $59335.6251_{-0.0001}^{+0.0001}$ & $59334.7092_{-0.0005}^{+0.0005}$ & $59357.697_{-0.010}^{+0.011}$ & \\
  $\tau$ [BJD - 2400000]& $59333.702_{-0.085}^{+0.090}$ & $59334.500_{-0.210}^{+0.146}$ & $59133.47_{-0.48}^{+0.55}$ &  \\
  $\Omega$ [deg] & $0.0$ & $-0.13_{-0.75}^{+0.72}$ & $-0.78_{-0.46}^{+0.35}$ & \\
  $(i_\mathrm{mut})_A-$ [deg] & $0.0$ & $1.17_{-0.78}^{+1.06}$ & $1.61_{-0.52}^{+0.40}$ & \\
  $(i_\mathrm{mut})_B-$ [deg] & $1.17_{-0.78}^{+1.06}$ & $0.0$ & $0.98_{-0.48}^{+0.56}$ & \\
  $\varpi^\mathrm{dyn}$ [deg]& $114_{-16}^{+14}$ & $72_{-38}^{+35}$ & $93.5_{-3.8}^{+3.8}$ & \\
  $i^\mathrm{dyn}$ [deg] & $1.45_{-0.48}^{+0.36}$ & $0.90_{-0.45}^{+0.47}$ & $0.16_{-0.04}^{+0.07}$ & \\
  $\Omega^\mathrm{dyn}$ [deg] & $329_{-25}^{+14}$ & $283_{-65}^{+65}$ & $139_{-21}^{+21}$ & \\
  $i_\mathrm{inv}$ [deg] & \multicolumn{3}{c}{$88.80_{-0.07}^{+0.07}$} & \\
  $\Omega_\mathrm{inv}$ [deg] & \multicolumn{3}{c}{$-0.69_{-0.42}^{+0.32}$} & \\
  \hline
  mass ratio $[q=M_\mathrm{sec}/M_\mathrm{pri}]$ & $0.918_{-0.009}^{+0.010}$ & $0.597_{-0.009}^{+0.008}$ & $0.806_{-0.004}^{+0.004}$ & \\
  $K^\mathrm{calc}_\mathrm{pri}$ [km\,s$^{-1}$] & $131.8_{-1.9}^{+1.9}$ & $108.9_{-1.6}^{+1.7}$ & $40.12_{-0.47}^{+0.56}$ & \\ 
  $K^\mathrm{calc}_\mathrm{sec}$ [km\,s$^{-1}$] & $143.4_{-1.3}^{+2.5}$ & $182.4_{-2.0}^{+2.3}$ & $49.79_{-0.71}^{+0.82}$ & \\ 
  \hline
  \multicolumn{5}{c}{Apsidal and nodal motion related parameters} \\
  \hline
$P_\mathrm{apse}$ [year] & $9.81_{-0.41}^{+0.40}$ & $3.99_{-0.23}^{+0.26}$ & $605_{-16}^{+16}$ &  \\ 
$P_\mathrm{apse}^\mathrm{dyn}$ [year] & $8.82_{-0.33}^{+0.33}$ & $3.84_{-0.21}^{+0.24}$ & $90.8_{-1.9}^{+1.9}$ & \\ 
$\Delta\omega_\mathrm{3b}$ [arcsec/cycle] & $161.6_{-3.2}^{+3.3}$ & $91.3_{-1.8}^{+1.8}$ &$8791_{-180}^{+182}$ &  \\ 
$\Delta\omega_\mathrm{GR}$ [arcsec/cycle] & $3.286_{-0.073}^{+0.088}$ & $3.673_{-0.073}^{+0.094}$ & $0.350_{-0.009}^{+0.011}$ & \\ 
$\Delta\omega_\mathrm{tide}$ [arcsec/cycle] & $669_{-29}^{+33}$ & $1209_{-77}^{+76}$ & $-$ & \\ 
  \hline  
\multicolumn{5}{c}{Stellar parameters} \\
\hline
   & Aa & Ab &  Ba & Bb  \\
  \hline
 \multicolumn{5}{c}{Relative quantities} \\
  \hline
 fractional radius [$R/a$]  & $0.2601_{-0.0025}^{+0.0023}$ & $0.2069_{-0.0053}^{+0.0062}$ & $0.3266_{-0.0047}^{+0.0046}$ & $0.1653_{-0.0046}^{+0.0053}$  \\
 temperature relative to $(T_\mathrm{eff})_\mathrm{Aa}$ & $1$ & $0.9989_{-0.0083}^{+0.0072}$ & $1.0049_{-0.0044}^{+0.0041}$ & $0.7545_{-0.0081}^{+0.0073}$  \\
 fractional flux [in TESS-band] & $0.3271_{-0.0143}^{+0.0139}$ & $0.2071_{-0.0037}^{+0.0046}$ & $0.2639_{-0.0108}^{+0.0119}$ & $0.0326_{-0.0013}^{+0.0015}$  \\
 fractional flux [in Cousins $R_C$-band] & $0.3399_{-0.0254}^{+0.0168}$ & $0.2137_{-0.0113}^{+0.0131}$ & $0.2638_{-0.0210}^{+0.0117}$ & $0.0290_{-0.0022}^{+0.0022}$ \\
 \hline
 \multicolumn{5}{c}{Physical Quantities} \\
  \hline 
 $M$ [M$_\odot$] & $2.333_{-0.067}^{+0.106}$ & $2.150_{-0.081}^{+0.083}$ & $2.261_{-0.068}^{+0.087}$ & $1.348_{-0.043}^{+0.055}$ \\
 $R$ [R$_\odot$] & $2.932_{-0.040}^{+0.052}$ & $2.333_{-0.080}^{+0.098}$ & $2.657_{-0.054}^{+0.053}$ & $1.343_{-0.049}^{+0.060}$ \\
 $T_\mathrm{eff}$ [K]& $8826_{-142}^{+171}$ & $8802_{-110}^{+186}$ & $8864_{-122}^{+181}$ & $6693_{-74}^{+91}$ \\
 $L_\mathrm{bol}$ [L$_\odot$] & $47.2_{-4.1}^{+4.0}$ & $29.7_{-3.3}^{+3.8}$ & $39.2_{-3.1}^{+4.0}$ & $3.26_{-0.34}^{+0.43}$ \\
 $M_\mathrm{bol}$ & $0.59_{-0.09}^{+0.10}$ & $1.09_{-0.13}^{+0.13}$ & $0.79_{-0.11}^{+0.09}$ & $3.49_{-0.13}^{+0.12}$ \\
 $M_V           $ & $0.60_{-0.07}^{+0.08}$ & $1.10_{-0.11}^{+0.12}$ & $0.81_{-0.08}^{+0.08}$ & $3.46_{-0.14}^{+0.13}$ \\
 $\log g$ [dex] & $3.871_{-0.009}^{+0.008}$ & $4.031_{-0.018}^{+0.017}$ & $3.943_{-0.011}^{+0.012}$ & $4.310_{-0.020}^{+0.019}$ \\
 \hline
\multicolumn{5}{c}{Global system parameters} \\
  \hline
$\log$(age) [dex] &\multicolumn{4}{c}{$8.764_{-0.053}^{+0.035}$} \\
$[M/H]$  [dex]    &\multicolumn{4}{c}{$0.079_{-0.087}^{+0.072}$} \\
$E(B-V)$ [mag]    &\multicolumn{4}{c}{$1.116_{-0.017}^{+0.018}$} \\
extra light $\ell_x$ [in TESS-band] & \multicolumn{4}{c}{$0.169_{-0.023}^{+0.022}$} \\
extra light $\ell_x$ [in Cousins $R_C$-band] & \multicolumn{4}{c}{$0.151_{-0.030}^{+0.056}$} \\
$(M_V)_\mathrm{tot}$  &\multicolumn{4}{c}{$-0.41_{-0.08}^{+0.09}$} \\
distance [pc]           & \multicolumn{4}{c}{$1464_{-36}^{+42}$} \\  
\hline
\end{tabular}}
\tablefoot{The RV amplitudes, denoted by `K' are calculated from the solutions, rather than fitted values, since no RV observations are available.}
\end{table*}

The results of our photodynamical fitting are given in Table~\ref{tab:syntheticfit_TIC433545934AB} in terms of the a-posteriori median parameter values, as well as their statistical 1$-\sigma$ uncertainties. (Note, the meanings of the symbols used in that table are given in Table~\ref{tbl:definitions}.) Moreover, the best-fit\footnote{By `best-fit' solution we mean the photodynamical solution where the total $\chi^2$, that is, $\chi^2=\chi^2_\mathrm{LC}+\chi^2_\mathrm{ETV}+\chi^2_\mathrm{SED}$ has its smallest value.} light curve of the outer eclipse, which was observed with TESS during Sector 38, is superposed on the data points in panel (a) of Fig.~\ref{fig:433545934Aoutereclipse}. The ETV curves of the two binaries, together with the best-fit photodynamical ETV curves are shown in the two panels of Fig.~\ref{fig:433545934ABETV}.

The best-fit model binary light curves are shown in (Fig.~\ref{fig:433545934ABlcfold}) as black curves.  These light curves are folded, or phased, to the orbital period.  The model phased curves are made from the best-fit photodynamical-solution light curves in a similar way to how the observed disentangled phase curves were produced.

In these plots one can see that the solutions nicely reproduce the long, flat, total eclipses in binary B and, moreover, they also yield  the feature that the primary eclipse of binary A is deeper by only a very small amount than the secondary eclipse of the same binary. Here, however, we run into an interesting issue. Zooming into the mid-eclipse areas of the folded binary A curve, one can see that these eclipses have a brief interval of totality. But, due to the photometric scatter in the original observations, it is unclear which eclipse is the transit and which is the occultation\footnote{When the smaller star passes in front of the disk of the larger one, this is referred to as the `transit' while, when the larger star totally eclipses the smaller one, this is called an `occultation'.}. When taking into account the practically circular orbit of binary A, it is clear that the star undergoing the primary eclipse (i.e., the deeper eclipse), has the slightly larger surface brightness (and, therefore, effective temperature) than its mate. In spite of this, it is not clear which star is the larger one. Our solutions clearly suggest, however, that the primary star (i.e., the hotter one is also the larger star. In all cases, when we exchanged the two stars in the fitting code, we found a substantially weaker solution. Therefore, for our final analysis we considered only those cases where the primary star was the larger one.

Taking an additional glance to the temperatures and temperature ratios in Table~\ref{tab:syntheticfit_TIC433545934AB}, we see that the resultant median effective temperature of binary A is $T_\mathrm{Aa}=8826^{+171}_{-142}$\,K and $T_\mathrm{Ab}=8802^{+186}_{-110}$\,K, while the median value of their ratio is $T_\mathrm{Ab}/T_\mathrm{Aa}=0.9989^{+0.0072}_{-0.0083}$. Therefore, both the medians of the absolute and the relative values suggest only a very slightly hotter primary. It is clear, however that the 1-$\sigma$ uncertainties allow that either of the two stars may be the hotter one, underlying the fact that their effective temperature differ by only a very small amount. On the other hand, the flat (or nearly flat) bottom of the eclipses of binary A show that the radii of the two stars differ substantially. Since we are considering main sequence (MS) stars, this situation can happen only near the end of the MS evolution when the stars begin to evolve up the giant branch. 

Here we also note that, considering both the absolute $T_{\rm eff}$ values and their ratios, we find that the hottest star of the entire quadruple system is the primary of binary B, as $T_\mathrm{Ba}=8864^{+181}_{-122}$\,K and $T_\mathrm{Ba}/T_\mathrm{Aa}=1.0049^{+0.0041}_{-0.0043}$ (while, again, the difference here is only weakly significant). This is so because the mass of star Ba was found to be between the mass of Aa and Ab and, therefore, this star is less evolved than star Aa.  

Considering the fourth star, i.e., Bb, even though it is clearly the least massive in the system, and its light contribution is by far the smallest, it is still more massive and hotter than our Sun. It is not clear, from an evolutionary point of view, why three of the stars in the system are so similar, while this fourth star, is the odd one out.

Here we comment on a few of the dynamical properties of the system.  If we compare the orbital elements of the outer orbit in Table~\ref{tab:syntheticfit_TIC433545934AB} with those that were obtained through the preliminary analytic study of the ETVs of binary A (given in Sect.~\ref{sec:periodstudy}), we can see that the analytic fit to the ETV fairly well approximated the results of the more accurate photodynamical analysis.  The outer orbit was found in the photodynamical analysis to be even more eccentric than was suggested by the analytic fit to the ETV, being $e_\mathrm{out}=0.623\pm0.006$. 

The orientation of the outer periastron, that is, $\omega_\mathrm{out}=274\degr\pm4\degr$, shows that the outer orbit is seen almost from the direction of pericenter, along its semi-major axis. Therefore, the observed extra eclipse, when binary B eclipsed binary A,  occurred at almost the outer pericenter. This fact, combined with the substantial eccentricity of the outer orbit, which is seen nearly, but not exactly, edge on ($i_\mathrm{out}=88\fdg91\pm0\fdg06$), explains why only the near-pericenter outer eclipses can be detected, while near-apocenter events (that is, when binary A would eclipse binary B) were not observed, as such events never occur.

Because the period of the outer orbit is more than two orders of magnitude longer than that of binary B, the orbital configuration of the quadruple is highly stable.  This is so even given the large eccentricity of the outer orbit.  Nonetheless, there are interesting dynamical effects, detectable in the ETV curves, that we were able to measure.  Figure \ref{fig:433545934ABETV} shows that the standard light travel time effects account for only about half of the ETV amplitude.  The rest is due to dynamical delays as each binary, in turn, slightly lengthens the orbital period of the other binary, and these changes vary considerably around the eccentric outer orbit. These dynamical delays are not only intrinsically interesting on an academic level, but they also directly contribute information that goes into unraveling the system parameters.

While numerous outer eclipses were detected in the archival data (see Sect.~\ref{sec:archive}) only a single outer eclipse was observed with TESS  near the periastron of the outer orbit (see the upper panel [a] of Fig.~\ref{fig:433545934Aoutereclipse}) In the lower panel of the same figure (b), we display the model total flux of binary A (that is, the flux contributions of stars Aa and Ab, without the flux contributions of stars Ba and Bb), as well as the individual fluxes of stars Aa and Ab. As one can see, star Aa, that is, the primary of binary A is eclipsed by the members of binary B during all three dips of the TESS outer eclipse event, while star Ab, that is, the secondary is eclipsed only during the middle event. Moreover, as is also mentioned in the Figure caption, during the secondary eclipses of binary A, the flux contribution of star Ab is exactly zero, since during the flat bottom of these secondary eclipses, the secondary component is totally eclipsed by its larger companion, that is the primary (Aa) of binary A.

\begin{figure}
\centering
     \includegraphics[width=\hsize]{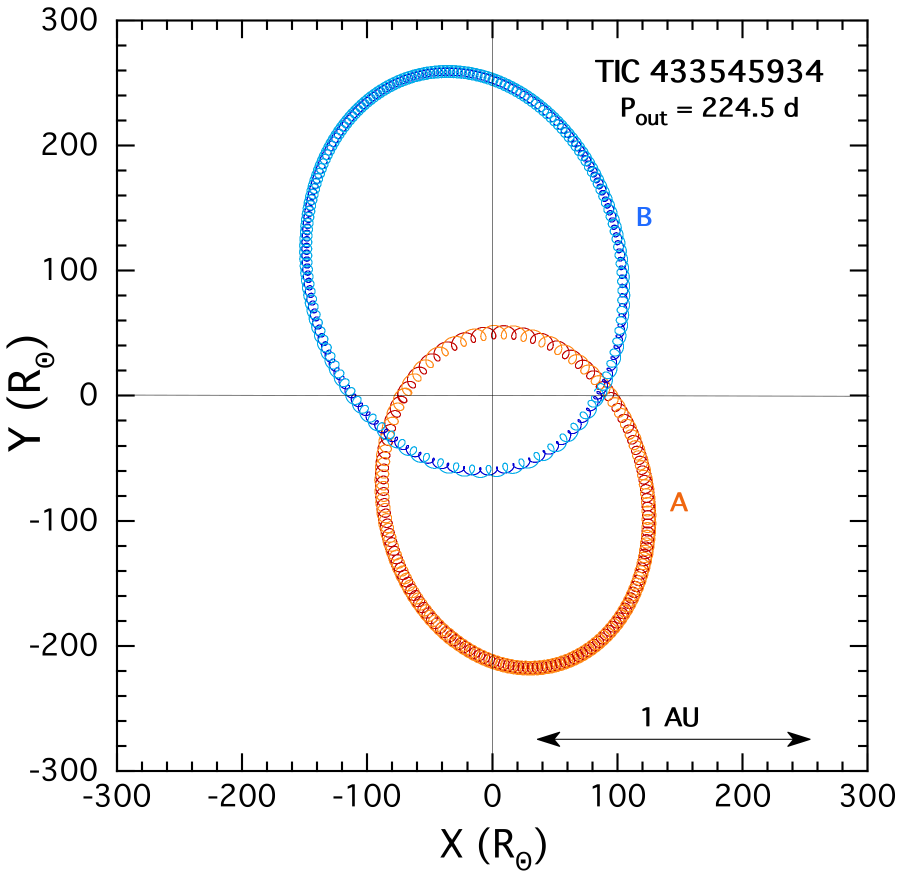}

     \caption{Top view of the orbits of all four stars in TIC 433545934. The three orbital planes are flat to within a couple of degrees.  The three orbital periods are 2.07, 1.41, and 224 days for the A, B and AB orbits, respectively, and the outer eccentricity is 0.6. The observer is in the $x-y$ plane at $y = - \infty$. Outer eclipses occur when the system is near periastron with the minimum separation between the two EBs.}
\label{fig:orbit_plot}
\end{figure}

Based on the derived system parameters, we show in Fig.~\ref{fig:orbit_plot} an overview of the entire TIC 433545934 system; specifically all three orbits are shown from a top view. This can be readily done because the system is practically flat, i.e., within $\sim$$2\degr$. Note, in this flatness we can see hints of the system's evolutionary origin. That is, we might infer that the system was formed from a single flat disk, with sequential disk collapse.

Finally, we comment briefly on the parameter uncertainties cited in Table \ref{tab:syntheticfit_TIC433545934AB}. These are statistical only and were generated without the use of RV data which were not available.  Because the \texttt{PARSEC} isochrones were utilized to partly compensate for the lack of RV information, one should also consider any systematic uncertainties in the stellar evolution models used to generate the \texttt{PARSEC} isochrones.  This affects mostly the absolute stellar parameters, particularly the stellar masses.  We note that the MCMC statistical errors listed in Table \ref{tab:syntheticfit_TIC433545934AB} for the masses, radii, $T_{\rm eff}$, and distance are about 3.5\%, 1.5-4\%, 1.5\%, and 2.7\%, respectively.  \citet{borkovitsetal22} explicitly examined in detail how well SED fitting works as a proxy for RVs in a system that had RV data.  In that work, they found that while masses could be determined to 1-3\% accuracy with RVs, the accuracy without the RVs was less at 3-10\%.  Thus, at least the uncertainties in the masses in Table \ref{tab:syntheticfit_TIC433545934AB} likely need to be somewhat inflated to be more realistic.  We note, however, that many of the adjusted parameters, especially non-dimensional quantities such as the fractional radii, mass ratios, and orbital elements, are practically independent of the masses and model-dependent systematics, and therefore the cited statistical errors are to be taken seriously\footnote{Regarding the photometric distance and its uncertainty, we note that in \citet{borkovitsetal25} we directly compared our photometric distances and uncertainties for 43 triply eclipsing triples with the \citet{bailer-jonesetal21} distances. Over the range of a few hundred to a few thousand pc, we found that the accuracy of our photometric distances was quite comparable to those of Gaia.  And, for most of these 43 triples, there were also no RV measurements available.}. 

\section{Summary, discussion, and conclusions}
\label{sec:discuss}

In this work we report the discovery and analysis of the first doubly eclipsing quadruple star system that also exhibits outer eclipses of one binary by the other -- TIC 433545934.  This outer eclipse is not just a novelty but, rather, the extra third-body eclipses provide vital information about the system parameters, some of which could not be obtained even with RV measurements.  In fact, without any RV measurements, we were able to extract all of the system parameters, including the detailed properties of all four stars, and the three sets of orbital elements. The stellar parameters are accurate at the 2\%-5\% level, the orbital angles ($i_{\rm mut}$ and the three inclination angles) are determined to better than a degree, and the outer eccentricity is good to better than 0.01. These accuracies were accomplished with a sophisticated photodynamical analysis that simultaneously models the photometric lightcurve, the eclipse timing variations, and the spectral energy distribution.  

Three of the four stars have masses that are between 2.2 and 2.4\,M$_\odot$, while the fourth star is just 30\% more massive than the Sun. All three of the more massive stars are substantially evolved away from the zero age main sequence, but are not yet completely through the main sequence.  The $T_{\rm eff}$ values of all three of the more massive stars are very close to 8800 K.  The system age is close to 580 Myr and the more massive stars in the A and B binaries will both overflow their Roche lobes in $\sim$152 Myr.  Thus, even though star Aa is the more massive, and currently more evolved, the Roche lobe of star Ba is smaller, and therefore the two primaries may actually overflow their respective Roche lobes at about the same time. While all three orbital planes are mutually well aligned to within $\lesssim 2^\circ$, i.e., the system is quite flat, the outer orbit is highly eccentric with $e_{\rm out} = 0.62$.  Note, these latter facts have substantial implications for the evolution of relatively tight outer orbit quadruples, however, such investigations are out of the scope of the current paper. 

We note that the outer period of this quadruple, $P_{\rm out} = 225$ d, implies a rather overall compact system; in particular, there are only four other known quadruples with shorter outer periods (ranging from 121-168 d; \citealp{pribullaetal23,powelletal25,kostovetal23}).  In fact, given that the shortest known outer period in a triple star system is $P_{\rm out} = 24.5$ days \citep{kostovetal24}, it will be interesting to see whether evolutionary pathways allow quadruples to form with outer periods between 24 d and 121 d.

While only a single outer eclipse from this quadruple star system was observed with TESS, we were able to utilize the ATLAS and ASAS-SN archival data to help nail down the outer period at 225 day, and to eliminate the possibility of comparably deep secondary outer eclipses. Without the archival data we might not have been completely certain of the outer period based entirely on the somewhat sparsely sampled ETV curves from the TESS data.

\begin{acknowledgements}
This project has received funding from the HUN-REN Hungarian Research Network.

T.\,B., T.\,M., I.\,B.\,B. acknowledge the financial support of the Hungarian National Research, Development and Innovation Office -- NKFIH Grants K-147131.
We would also like to thank the Pierre Auger Collaboration for the use of its facilities. The operation of the robotic telescope FRAM is supported by the grant of the Ministry of Education of the Czech Republic LM2018102. The data calibration and analysis related to the FRAM telescope is supported by the Ministry of Education of the Czech Republic MSMT-CR LTT18004, MSMT/EU funds \verb|CZ.02.1.01/0.0/0.0/16_013/0001402| and \verb|CZ.02.1.01/0.0/0.0/18_046/0016010|. 

We are also grateful to the ESO team at the La Silla Observatory for their help in maintaining and operating the Danish 1.54m telescope. The research of P.Z. was also supported by the project {\sc Cooperatio - Physics} of Charles University in Prague.

V. B. K. acknowledges support from NASA grant 80NSSC23K0270.

This paper makes extensive use of data collected by the TESS mission. Funding for the TESS mission is provided by the NASA Science Mission directorate. Some of the data presented in this paper were obtained from the Mikulski Archive for Space Telescopes (MAST). STScI is operated by the Association of Universities for Research in Astronomy, Inc., under NASA contract NAS5-26555. Support for MAST for non-HST data is provided by the NASA Office of Space Science via grant NNX09AF08G and by other grants and contracts.

Distances and other astrometric properties for the target  were taken from the archives of European  Space Agency (ESA)  mission \textit{Gaia}\footnote{\url{https://www.cosmos.esa.int/gaia}},  processed  by  the \textit{Gaia}   Data   Processing   and  Analysis   Consortium   (DPAC)\footnote{\url{https://www.cosmos.esa.int/web/gaia/dpac/consortium}}.  Funding for the DPAC  has been provided  by national  institutions, in  particular the institutions participating in the \textit{Gaia} Multilateral Agreement.

Some of the SED fluxes and magnitudes were obtained with the Wide-field Infrared Survey Explorer, which is a joint project of the University of California, Los Angeles, and the Jet Propulsion Laboratory/California Institute of Technology, funded by the National Aeronautics and Space Administration. 

Additionally, some of the SED fluxes and magnitudes were obtained with the Two Micron All Sky Survey, which is a joint project of the University of Massachusetts and the Infrared Processing and Analysis Center/California Institute of Technology, funded by the National Aeronautics and Space Administration and the National Science Foundation.

We  used the  Simbad  service  operated by  the  Centre des  Donn\'ees Stellaires (Strasbourg,  France). 

This research has also made use of the VizieR catalogue access tool, CDS, Strasbourg, France (DOI : 10.26093/cds/vizier). The original description of the VizieR service was published in \citet{ochsenbein00}.

\end{acknowledgements}

\begin{appendix} 

\onecolumn

\section{Eclipse times of the two inner EBs of the 2+2 quadruple system TIC\,433545934}
\label{app:ToMs}

In this appendix we tabulate the times of the individual primary and secondary eclipses of the two inner EBs of TIC\,433545934.  These naturally include mostly eclipses from TESS, plus a few that were observed from the ground.

\begin{table*}[ht]
\caption{Times of minima of TIC 433545934A}
 \label{Tab:TIC_433545934A_ToM}
\scalebox{0.88}{\begin{tabular}{@{}lrllrllrllrl}
\hline
BJD & Cycle  & std. dev. & BJD & Cycle  & std. dev. & BJD & Cycle  & std. dev. & BJD & Cycle  & std. dev. \\ 
$-2\,400\,000$ & no. &   \multicolumn{1}{c}{$(d)$} & $-2\,400\,000$ & no. &   \multicolumn{1}{c}{$(d)$} & $-2\,400\,000$ & no. &   \multicolumn{1}{c}{$(d)$} & $-2\,400\,000$ & no. &   \multicolumn{1}{c}{$(d)$} \\ 
\hline
59334.58608 &    -0.5 & 0.00008 & 60050.64667 &   345.0 & 0.00007 & 60087.94766 &   363.0 & 0.00009 & 61085.85062 &   844.5 & 0.00008  \\ 
59335.62280 &     0.0 & 0.00010 & 60051.68299 &   345.5 & 0.00009 & 60090.02066 &   364.0 & 0.00008 & 61086.88643 &   845.0 & 0.00009  \\ 
59336.65863 &     0.5 & 0.00008 & 60052.71860 &   346.0 & 0.00008 & 60091.05540 &   364.5 & 0.00010 & 61088.95917 &   846.0 & 0.00006  \\ 
59337.69610 &     1.0 & 0.00012 & 60053.75576 &   346.5 & 0.00008 & 60092.09158 &   365.0 & 0.00008 & 61089.99530 &   846.5 & 0.00006  \\ 
59338.73307 &     1.5 & 0.00014 & 60055.82809 &   347.5 & 0.00008 & 60094.16420 &   366.0 & 0.00010 & 61091.03159 &   847.0 & 0.00009  \\ 
59339.76785 &     2.0 & 0.00007 & 60056.86322 &   348.0 & 0.00008 & 60095.20014 &   366.5 & 0.00010 & 61092.06762 &   847.5 & 0.00007  \\ 
59340.80541 &     2.5 & 0.00011 & 60057.89948 &   348.5 & 0.00009 & 60096.23696 &   367.0 & 0.00011 & 61093.10355 &   848.0 & 0.00011  \\ 
59341.84110 &     3.0 & 0.00010 & 60058.93559 &   349.0 & 0.00007 & 61046.47491 &   825.5 & 0.00007 & 61095.17666 &   849.0 & 0.00006  \\ 
59342.87752 &     3.5 & 0.00008 & 60059.97248 &   349.5 & 0.00008 & 61047.51026 &   826.0 & 0.00007 & 61096.21263 &   849.5 & 0.00006  \\ 
59343.91426 &     4.0 & 0.00010 & 60061.00865 &   350.0 & 0.00009 & 61048.54715 &   826.5 & 0.00007 & 61097.24941 &   850.0 & 0.00007  \\ 
59344.95074 &     4.5 & 0.00009 & 60062.04408 &   350.5 & 0.00010 & 61049.58313 &   827.0 & 0.00007 & 61098.28541 &   850.5 & 0.00010  \\ 
59345.98782 &     5.0 & 0.00010 & 60063.07999 &   351.0 & 0.00008 & 61050.62054 &   827.5 & 0.00008 & 61099.32117 &   851.0 & 0.00009  \\ 
59348.06047 &     6.0 & 0.00008 & 60064.11621 &   351.5 & 0.00010 & 61051.65498 &   828.0 & 0.00008 & 61100.35867 &   851.5 & 0.00007  \\ 
59349.09630 &     6.5 & 0.00010 & 60065.15312 &   352.0 & 0.00006 & 61053.72750 &   829.0 & 0.00007 & 61101.39337 &   852.0 & 0.00012  \\ 
59350.13488 &     7.0 & 0.00008 & 60066.18838 &   352.5 & 0.00008 & 61054.76375 &   829.5 & 0.00007 & 61102.43018 &   852.5 & 0.00010  \\ 
59351.16960 &     7.5 & 0.00008 & 60067.22507 &   353.0 & 0.00008 & 61063.05183 &   833.5 & 0.00017 & 61103.46677 &   853.0 & 0.00012  \\ 
59352.20718 &     8.0 & 0.00007 & 60069.29667 &   354.0 & 0.00008 & 61064.08926 &   834.0 & 0.00007 & 61104.50271 &   853.5 & 0.00010  \\ 
59353.24258 &     8.5 & 0.00009 & 60070.33408 &   354.5 & 0.00010 & 61065.12552 &   834.5 & 0.00008 & 61105.53951 &   854.0 & 0.00012  \\ 
59354.28067 &     9.0 & 0.00009 & 60071.36900 &   355.0 & 0.00010 & 61066.16341 &   835.0 & 0.00007 & 61106.57582 &   854.5 & 0.00012  \\ 
59355.31660 &     9.5 & 0.00007 & 60072.40450 &   355.5 & 0.00009 & 61068.23445 &   836.0 & 0.00008 & 61107.61235 &   855.0 & 0.00011  \\ 
59356.35416 &    10.0 & 0.00008 & 60073.44082 &   356.0 & 0.00007 & 61069.27011 &   836.5 & 0.00007 & 61108.64768 &   855.5 & 0.00010  \\ 
59359.46489 &    11.5 & 0.00007 & 60074.47715 &   356.5 & 0.00010 & 61071.34332 &   837.5 & 0.00008 & 61109.68393 &   856.0 & 0.00010  \\ 
59360.50110 &    12.0 & 0.00012 & 60075.51394 &   357.0 & 0.00008 & 61072.37954 &   838.0 & 0.00008 & 61110.72032 &   856.5 & 0.00011  \\ 
59648.57826 &   151.0 & 0.00016 & 60076.55135 &   357.5 & 0.00009 & 61073.41481 &   838.5 & 0.00114 & 61111.75688 &   857.0 & 0.00010  \\ 
59650.64590 &   152.0 & 0.00007 & 60077.58575 &   358.0 & 0.00009 & 61074.45260 &   839.0 & 0.00006 & 61115.90201 &   859.0 & 0.00010  \\ 
59679.65642 &   166.0 & 0.00007 & 60078.62182 &   358.5 & 0.00008 & 61075.48669 &   839.5 & 0.00007 & 61116.93772 &   859.5 & 0.00010  \\ 
60041.32205 &   340.5 & 0.00009 & 60079.65949 &   359.0 & 0.00009 & 61076.52474 &   840.0 & 0.00006 & 61117.97538 &   860.0 & 0.00010  \\ 
60042.35782 &   341.0 & 0.00008 & 60080.69387 &   359.5 & 0.00009 & 61078.59688 &   841.0 & 0.00006 & 61119.01030 &   860.5 & 0.00010  \\ 
60043.39413 &   341.5 & 0.00008 & 60081.73113 &   360.0 & 0.00010 & 61079.63211 &   841.5 & 0.00008 & 61121.08471 &   861.5 & 0.00012  \\ 
60044.42954 &   342.0 & 0.00009 & 60082.76640 &   360.5 & 0.00010 & 61080.66943 &   842.0 & 0.00012 & 61122.12051 &   862.0 & 0.00009  \\ 
60045.46609 &   342.5 & 0.00009 & 60083.80193 &   361.0 & 0.00010 & 61081.70538 &   842.5 & 0.00007 & 61123.15561 &   862.5 & 0.00009  \\ 
60046.50232 &   343.0 & 0.00007 & 60084.83835 &   361.5 & 0.00008 & 61082.74146 &   843.0 & 0.00006 & 61124.19372 &   863.0 & 0.00010  \\ 
60047.53803 &   343.5 & 0.00008 & 60085.87478 &   362.0 & 0.00007 & 61083.77792 &   843.5 & 0.00006 & 61125.22931 &   863.5 & 0.00012  \\ 
60048.57410 &   344.0 & 0.00008 & 60086.91023 &   362.5 & 0.00009 & 61084.81412 &   844.0 & 0.00007 & 61126.26651 &   864.0 & 0.00026  \\ 
60049.61007 &   344.5 & 0.00008  \\ 
\hline
\end{tabular}}
\tablefoot{Integer and half-integer cycle numbers denote primary and secondary eclipses, respectively. Eclipse times related to cycle numbers 151.0, 152.0 and 166.0 are calculated from ground-based follow up observations. All the other mid-minima times are dervived from TESS light curves.} 
\end{table*}

\begin{table*}
\caption{Times of minima of TIC 433545934B}
 \label{Tab:TIC_433545934B_ToM}
\scalebox{0.88}{\begin{tabular}{@{}lrllrllrllrl}
\hline
BJD & Cycle  & std. dev. & BJD & Cycle  & std. dev. & BJD & Cycle  & std. dev. & BJD & Cycle  & std. dev. \\ 
$-2\,400\,000$ & no. &   \multicolumn{1}{c}{$(d)$} & $-2\,400\,000$ & no. &   \multicolumn{1}{c}{$(d)$} & $-2\,400\,000$ & no. &   \multicolumn{1}{c}{$(d)$} & $-2\,400\,000$ & no. &   \multicolumn{1}{c}{$(d)$} \\ 
\hline
59334.00543 &    -0.5 & 0.00047 & 60053.56606 &   509.0 & 0.00018 & 60093.82239 &   537.5 & 0.00050 & 61090.89019 &  1243.5 & 0.00031  \\ 
59334.71048 &     0.0 & 0.00030 & 60054.28125 &   509.5 & 0.00042 & 60094.52644 &   538.0 & 0.00023 & 61091.58590 &  1244.0 & 0.00016  \\ 
59335.41459 &     0.5 & 0.00041 & 60054.97955 &   510.0 & 0.00023 & 60095.93855 &   539.0 & 0.00026 & 61092.29574 &  1244.5 & 0.00034  \\ 
59336.12272 &     1.0 & 0.00030 & 60055.68486 &   510.5 & 0.00038 & 61046.39671 &  1212.0 & 0.00022 & 61092.99860 &  1245.0 & 0.00025  \\ 
59336.83115 &     1.5 & 0.00050 & 60056.39466 &   511.0 & 0.00017 & 61047.10835 &  1212.5 & 0.00042 & 61093.70779 &  1245.5 & 0.00079  \\ 
59337.53323 &     2.0 & 0.00023 & 60057.10114 &   511.5 & 0.00036 & 61047.80937 &  1213.0 & 0.00020 & 61094.40940 &  1246.0 & 0.00018  \\ 
59338.24490 &     2.5 & 0.00034 & 60057.80370 &   512.0 & 0.00016 & 61048.52204 &  1213.5 & 0.00034 & 61095.82248 &  1247.0 & 0.00019  \\ 
59338.94406 &     3.0 & 0.00041 & 60058.51179 &   512.5 & 0.00047 & 61049.22236 &  1214.0 & 0.00021 & 61096.52982 &  1247.5 & 0.00029  \\ 
59339.65554 &     3.5 & 0.00064 & 60059.21774 &   513.0 & 0.00016 & 61049.93024 &  1214.5 & 0.00034 & 61097.23341 &  1248.0 & 0.00021  \\ 
59340.35926 &     4.0 & 0.00019 & 60059.92473 &   513.5 & 0.00042 & 61050.63777 &  1215.0 & 0.00019 & 61097.94306 &  1248.5 & 0.00044  \\ 
59341.06438 &     4.5 & 0.00049 & 60060.63052 &   514.0 & 0.00022 & 61051.34459 &  1215.5 & 0.00029 & 61098.64750 &  1249.0 & 0.00016  \\ 
59341.76859 &     5.0 & 0.00027 & 60061.34017 &   514.5 & 0.00048 & 61052.04502 &  1216.0 & 0.00018 & 61099.35757 &  1249.5 & 0.00043  \\ 
59342.47521 &     5.5 & 0.00073 & 60062.04328 &   515.0 & 0.00023 & 61053.45996 &  1217.0 & 0.00019 & 61100.06140 &  1250.0 & 0.00024  \\ 
59343.18311 &     6.0 & 0.00027 & 60062.74423 &   515.5 & 0.00033 & 61054.16784 &  1217.5 & 0.00031 & 61100.77220 &  1250.5 & 0.00044  \\ 
59343.88897 &     6.5 & 0.00068 & 60063.45622 &   516.0 & 0.00020 & 61054.86815 &  1218.0 & 0.00018 & 61101.47653 &  1251.0 & 0.00029  \\ 
59344.59335 &     7.0 & 0.00031 & 60064.16805 &   516.5 & 0.00043 & 61055.57890 &  1218.5 & 0.00030 & 61102.18153 &  1251.5 & 0.00052  \\ 
59345.30615 &     7.5 & 0.00070 & 60064.86655 &   517.0 & 0.00017 & 61063.34504 &  1224.0 & 0.00016 & 61102.88141 &  1252.0 & 0.00023  \\ 
59346.00545 &     8.0 & 0.00023 & 60065.57237 &   517.5 & 0.00055 & 61064.05367 &  1224.5 & 0.00030 & 61103.59846 &  1252.5 & 0.00053  \\ 
59347.42047 &     9.0 & 0.00027 & 60066.28149 &   518.0 & 0.00017 & 61064.75631 &  1225.0 & 0.00018 & 61104.29322 &  1253.0 & 0.00028  \\ 
59348.13219 &     9.5 & 0.00077 & 60066.98879 &   518.5 & 0.00047 & 61065.46414 &  1225.5 & 0.00041 & 61105.00538 &  1253.5 & 0.00046  \\ 
59348.83666 &    10.0 & 0.00056 & 60067.69412 &   519.0 & 0.00023 & 61066.16856 &  1226.0 & 0.00020 & 61105.70613 &  1254.0 & 0.00030  \\ 
59349.54481 &    10.5 & 0.00066 & 60069.10418 &   520.0 & 0.00023 & 61066.87950 &  1226.5 & 0.00033 & 61106.41219 &  1254.5 & 0.00048  \\ 
59350.24740 &    11.0 & 0.00031 & 60069.81599 &   520.5 & 0.00035 & 61067.58043 &  1227.0 & 0.00016 & 61107.82953 &  1255.5 & 0.00060  \\ 
59351.65910 &    12.0 & 0.00025 & 60070.51642 &   521.0 & 0.00018 & 61068.99241 &  1228.0 & 0.00018 & 61108.52996 &  1256.0 & 0.00022  \\ 
59352.37042 &    12.5 & 0.00042 & 60071.23180 &   521.5 & 0.00035 & 61069.70237 &  1228.5 & 0.00028 & 61109.23848 &  1256.5 & 0.00054  \\ 
59353.07206 &    13.0 & 0.00037 & 60071.93174 &   522.0 & 0.00024 & 61070.41118 &  1229.0 & 0.00017 & 61109.94294 &  1257.0 & 0.00028  \\ 
59353.78288 &    13.5 & 0.00061 & 60072.63960 &   522.5 & 0.00036 & 61071.11105 &  1229.5 & 0.00044 & 61110.65200 &  1257.5 & 0.00046  \\ 
59354.48275 &    14.0 & 0.00022 & 60073.34595 &   523.0 & 0.00018 & 61071.81569 &  1230.0 & 0.00022 & 61111.35529 &  1258.0 & 0.00022  \\ 
59355.19162 &    14.5 & 0.00050 & 60074.04844 &   523.5 & 0.00041 & 61073.22987 &  1231.0 & 0.00049 & 61112.06364 &  1258.5 & 0.00050  \\ 
59355.89391 &    15.0 & 0.00024 & 60074.75420 &   524.0 & 0.00020 & 61074.63723 &  1232.0 & 0.00015 & 61112.76609 &  1259.0 & 0.00025  \\ 
59357.30383 &    16.0 & 0.00065 & 60075.46818 &   524.5 & 0.00035 & 61076.05255 &  1233.0 & 0.00019 & 61113.47732 &  1259.5 & 0.00140  \\ 
59358.71733 &    17.0 & 0.00035 & 60076.86920 &   525.5 & 0.00033 & 61076.75917 &  1233.5 & 0.00027 & 61114.17834 &  1260.0 & 0.00022  \\ 
59360.13308 &    18.0 & 0.00028 & 60077.57668 &   526.0 & 0.00018 & 61077.47494 &  1234.0 & 0.00020 & 61114.88775 &  1260.5 & 0.00050  \\ 
60041.56390 &   500.5 & 0.00042 & 60078.28228 &   526.5 & 0.00042 & 61078.17537 &  1234.5 & 0.00034 & 61115.59226 &  1261.0 & 0.00026  \\ 
60042.26264 &   501.0 & 0.00022 & 60078.98851 &   527.0 & 0.00019 & 61078.87721 &  1235.0 & 0.00021 & 61116.30401 &  1261.5 & 0.00040  \\ 
60042.97479 &   501.5 & 0.00039 & 60080.40172 &   528.0 & 0.00028 & 61080.28855 &  1236.0 & 0.00016 & 61117.00625 &  1262.0 & 0.00028  \\ 
60043.67781 &   502.0 & 0.00018 & 60081.11087 &   528.5 & 0.00042 & 61080.99486 &  1236.5 & 0.00036 & 61117.71558 &  1262.5 & 0.00060  \\ 
60045.09100 &   503.0 & 0.00025 & 60082.52039 &   529.5 & 0.00046 & 61081.70168 &  1237.0 & 0.00017 & 61118.41710 &  1263.0 & 0.00026  \\ 
60045.79676 &   503.5 & 0.00037 & 60083.22876 &   530.0 & 0.00025 & 61082.41144 &  1237.5 & 0.00038 & 61119.12885 &  1263.5 & 0.00049  \\ 
60046.51207 &   504.0 & 0.00016 & 60083.93796 &   530.5 & 0.00036 & 61083.11392 &  1238.0 & 0.00019 & 61119.82811 &  1264.0 & 0.00035  \\ 
60047.21028 &   504.5 & 0.00045 & 60084.64247 &   531.0 & 0.00025 & 61083.81609 &  1238.5 & 0.00031 & 61120.53703 &  1264.5 & 0.00050  \\ 
60047.91356 &   505.0 & 0.00016 & 60086.05412 &   532.0 & 0.00016 & 61084.52568 &  1239.0 & 0.00016 & 61121.23894 &  1265.0 & 0.00028  \\ 
60048.63300 &   505.5 & 0.00037 & 60087.46739 &   533.0 & 0.00021 & 61085.23423 &  1239.5 & 0.00037 & 61121.94547 &  1265.5 & 0.00055  \\ 
60049.32659 &   506.0 & 0.00025 & 60088.17609 &   533.5 & 0.00054 & 61085.93687 &  1240.0 & 0.00019 & 61122.65189 &  1266.0 & 0.00022  \\ 
60050.03740 &   506.5 & 0.00042 & 60088.88431 &   534.0 & 0.00019 & 61086.64958 &  1240.5 & 0.00047 & 61123.36261 &  1266.5 & 0.00063  \\ 
60050.74052 &   507.0 & 0.00014 & 60089.58334 &   534.5 & 0.00041 & 61088.76357 &  1242.0 & 0.00016 & 61124.06274 &  1267.0 & 0.00025  \\ 
60051.44800 &   507.5 & 0.00046 & 60091.00165 &   535.5 & 0.00024 & 61089.47068 &  1242.5 & 0.00028 & 61124.77186 &  1267.5 & 0.00054  \\ 
60052.15341 &   508.0 & 0.00017 & 60091.70147 &   536.0 & 0.00022 & 61090.17175 &  1243.0 & 0.00016 & 61125.47370 &  1268.0 & 0.00038  \\ 
60052.85774 &   508.5 & 0.00038 & 60092.41291 &   536.5 & 0.00037  \\ 
\hline
\end{tabular}}
\tablefoot{Integer and half-integer cycle numbers denote primary and secondary eclipses, respectively.} 
\end{table*}

\end{appendix}

\end{document}